\documentclass[landscape,usenatbib,usegraphicx]{mn2e}
\newcommand{\be}{\begin{eqnarray}}
\newcommand{\ee}{\end{eqnarray}}

\def\theequation{\thesection.\arabic{equation}}
\title[Dipole straylight contamination in CMB anisotropy missions
]{On the dipole straylight contamination in spinning space missions dedicated to CMB anisotropy}

\author[C.Burigana, A.Gruppuso and F.Finelli]
{C.~Burigana \thanks{burigana@iasfbo.inaf.it},$^1$ 
A.~Gruppuso \thanks{gruppuso@iasfbo.inaf.it}$^1$ and
F.~Finelli \thanks{finelli@iasfbo.inaf.it}$^{1,2}$
\\
$^1$ INAF-IASF Bologna, Istituto di Astrofisica Spaziale e Fisica Cosmica di Bologna \\
Istituto Nazionale di Astrofisica, via Gobetti 101, I-40129 Bologna - Italy \\
$^2$ INAF-OAB, Osservatorio Astronomico di Bologna \\
Istituto Nazionale di Astrofisica,
via Ranzani 1, I-40127 Bologna - Italy
}

\begin{document}

\def\lsim{\,\lower2truept\hbox{${< \atop\hbox{\raise4truept\hbox{$\sim$}}}$}\,}
\def\gsim{\,\lower2truept\hbox{${> \atop\hbox{\raise4truept\hbox{$\sim$}}}$}\,}

\maketitle

\begin{abstract}

We present an analysis of the dipole straylight contamination (DSC) for spinning
space-missions designed to measure CMB 
anisotropies. Although this work is mainly devoted to the {\sc Planck} project,
it is relatively general and allows to focus on
the most relevant DSC implications. 
We first study a simple 
analytical model for the DSC in which 
the pointing direction of the main spillover can be assumed 
parallel or not to the spacecraft spin axis direction and 
compute the time ordered data and map. 
The map is then analysed paying particular attention to the 
DSC of the low multipole coefficients of the map. 
Through dedicated numerical simulations
we verify the analytical results and extend 
the analysis to higher multipoles
and to more complex (and realistic) cases 
by relaxing some of the simple assumptions adopted in the 
analytical approach.
We find that the systematic effect averages out in an even number of surveys, except 
for a contamination of the dipole itself that survives when spin axis and spillover 
directions are not parallel and for a contamination of the other multipoles
in the case of complex scanning strategies.
In particular, the observed quadrupole 
can be affected by the DSC 
in an odd number of surveys or in the presence of survey uncompleteness or
over-completeness.
Various aspects relevant in CMB space projects
(such as implications for calibration, impact on polarization measurements, accuracy requirement
in the far beam knowledge for data analysis applications, 
scanning strategy dependence) are discussed.

\end{abstract}

\begin{keywords}
Cosmology: cosmic microwave background -- space vehicles: instruments --
   methods: data analysis.
\end{keywords}

\raggedbottom
\setcounter{page}{1}
\section{Introduction}
\setcounter{equation}{0}
\label{intro}
The observed pattern of the cosmic microwave background (CMB) anisotropies is 
dominated by the dipole signal originating from the motion of the Sun
in the CMB rest frame. This signal is two orders of 
magnitude larger than the typical cosmological effect. While the dipole subtraction is of primary 
importance to map CMB anisotropies, 
little is known about the straylight contamination induced by this dipole pattern.
The straylight contamination 
consists in unwanted radiation entering the beam at large
angles from the antenna boresight direction.
The most relevant contribution is expected from a particular far beam extended region,
the so-called main spillover, typically located at $\sim 40^\circ-90^\circ$
from the main optical axis. 
In the case of the {\sc Planck}~\footnote{http://www.rssd.esa.int/planck}
telescope, its peak response is at $\sim 90^\circ$ from the telescope optical axis
\citep{sandri} and points far away from the Sun along directions not far from 
the spacecraft spin axis.
It is already known that the straylight contamination of other non-cosmological effects, 
such as Galactic emissions \citep{burigana1,burigana2,sandri}
and inner Solar System bodies \citep{burigana_straysun,burigana_beamrec}, 
plays an important role in designing 
space-missions like {\sc Planck}
\citep{lfi,mandolesi,hfi,lamarre,tauber,tauber06}. Its control and removal represents also 
a relevant aspect of WMAP~\footnote{http://lambda.gsfc.nasa.gov/product/map/current/}
design and data analysis \citep{barnes,jarosik06}.

The aim of this paper is to study the dipole straylight contamination 
(DSC) with analytical and numerical methods.
We start building a model sufficiently simple to be treated analytically 
that allows to understand to first order how the straylight features affect 
the recoverd CMB angular power spectrum in a way largely independent of the specific optical and
scanning strategy details. 
The analysis is first performed considering the direction of pointing of the main spillover centre 
parallel or not to the spin axis but keeping the the main spillover centre
direction in the
plane defined by the telescope and spin axis directions.
In this work, we call $\alpha$ the angle between the main spillover centre
direction and the spin axis direction.
We then extend the analysis allowing for displacements of the 
main spillover centre direction from the
plane defined by the telescope axis and the spin axis.
The analytical results are presented perturbatively in $\alpha$.

We shall work in the fixed frame with the satellite with axes
pointing fixed (far away) stars. In this frame the vector
associated to the dipole is constant while the main spillover direction is not
(it rotates of $2 \pi$ in $1$ year). We consider the dipole for the
motion of the Sun with respect to the rest frame of the CMB and we
neglect, for simplicity, small deviations due to the motion of the
Earth around the Sun.

We implement dedicated numerical simulations
in order to check the analytical results and extend 
the analysis to more complex (and realistic) cases 
by relaxing some of the simple assumptions adopted in the 
analytical approach.
In particular, simulations allow to consider
non-small values of the angle $\alpha$ and complicate scanning strategies
introducing the effect associated to the survey uncompleteness
(or over-completeness)
and extending the analysis to higher multipoles, $\ell$.

Finally we discuss how the DSC may affect the multipole pattern, with 
particular emphasis paid to the possible connection with the small signal 
observed by COBE/DMR \citep{COBE} and WMAP \citep{bennett,spergel03,spergel06} 
at low $\ell$.

The article is organized as follows. In Section \ref{dipole} the
convolution of the dipole and the straylight beam is presented. 
The analytical model for the beam response in the main spillover region is
presented in Section \ref{model} where the map is computed with 
particular care to low multipoles (up to $\ell=4$). 
In Section \ref{implicationforcosmology} the power spectrum is discussed and the 
DSC of the quadrupole ($\ell =2$) is analysed. In Section \ref{numericalsimulations} 
numerical simulations are presented and discussed. 
In Section~\ref{displacement} we extend the analysis to arbitrary 
directions of the main spillover. The main implications for 
spinning space missions dedicated to the CMB anisotropy
are discussed in Section~\ref{numremarks}.
In Section \ref{statistic}
a statistical analysis for the amplitude of the quadrupole 
is presented.
Finally, our main conclusions are drawn in Section \ref{conclusion}.

\section{The dipole and the straylight beam}
\setcounter{equation}{0}
\label{dipole}

We start considering the convolution $I$ of the dipole with the
main spillover: 
\be I = \int d\Omega \, T_{1 m} Y_1^m (\theta,
\varphi) B_{SL}(\theta, \varphi) \, , \label{conv}
\ee
where
$d\Omega $ is the element of solid angle, $d\Omega = d \theta \sin
\theta \, d \varphi $ with the colatitude $\theta \in \left[0,\pi \right]$ and
the longitude $\varphi \in \left[0, 2 \pi\right.\left[\right.$, the sum on $m$
over $-1,0,1$ is understood, $T_{1 m}$ are the coefficients of the
expansion of the dipole \footnote{We use the symbol $T_{\ell m}$
because we want to make clear that the dimensionality is given
by a temperature (e.g. in $\mu$K).
} on the basis of the spherical harmonics $Y_1^m
(\theta, \varphi)$, and $B_{SL}(\theta, \varphi)$ is the beam response
representing the shape of the main spillover in the $(\theta,
\varphi)$-plane. In this notation $B_{SL}$ is normalized to the
whole beam integrated response, dominated by the contribution
in the main beam 
\be
\int_{4\pi} d\Omega B \simeq \int_{{\rm main~beam}} d\Omega B \simeq 2\pi \sigma_b^2
\ee
where $\sigma_b={\rm FWHM}/\sqrt{8 \,\ln 2}$,
with ${\rm FWHM}$ \footnote{FWHM = Full Width Half Maximum.}
representing the main beam angular resolution.

The convolution $I$ can be rewritten in the following way: 
\begin{eqnarray} I =
\sqrt{3 \over {4 \pi }} \left[  T_{10} \int d\theta \, d \varphi
\, \sin \theta \, \cos \theta \, B_{SL}(\theta, \varphi) - \right. \nonumber \\
\left. \sqrt{2} \, \int d\theta \, d \varphi \, \sin ^2 \theta \, Re
\left[ T_{11} e^{i \varphi } \right] \, B_{SL}(\theta, \varphi)
\right] \,  \label{conv2} 
\end{eqnarray} 
where $Re\left[...\right]$ is the real part of $\left[...\right]$. 
In order to obtain equation~(\ref{conv2}) it has been used that 
$ T_{1-1}= - T^{\star}_{11}$, where the symbol $^{\star}$ means complex conjugation.
For sake of completeness, we write the spherical harmonics for $\ell=1$ \citep{sakurai}:
\begin{eqnarray} 
& & Y_1^0 (\theta, \varphi) = \sqrt{3 \over {4 \pi }} \cos \theta
\, , \\
& & Y_1^{\pm 1} (\theta, \varphi) = \mp \sqrt{3 \over {8 \pi }} 
e^{\pm i \varphi } \sin \theta \, .
\end{eqnarray}
Although it may not be clear from the notation of equation~(\ref{conv2}), 
notice that $I$ is a function of the geometric features of the shape of
the main spillover in the $(\theta,\varphi)$-plane.

\section{Analytical models}
\setcounter{equation}{0}
\label{model}

Equation~(\ref{conv2}) is general and exact. 
Any specific approximation of the beam response 
$B_{SL}$ inspired by experiments 
will introduce a certain degree of uncertainty with respect to 
the real case. 
We introduce two different parametrizations for $B_{SL}$:
we first consider the {\it Top Hat} approximation, analysed in Subsection \ref{tophat}, 
and then the {\it Gaussian} approximation, studied in Subsection
\ref{gaussian}. 
We shall see in the following that the two descriptions are quite similar 
and are formally equivalent when the main spillover area is sufficiently small
(see Subsection \ref{link}).
For this reason we concentrate on the {\it Top Hat} case
for the analytical estimate of the DSC. 

\subsection{Top hat approximation}
\label{tophat}

Our first approximation \citep{report2005}
for $B_{SL}$ is the {\it Top Hat}:
\begin{eqnarray} B_{SL} (\theta, \varphi) = \, f_{SL}
\, \Delta(\, \theta ,\, \theta_{ms} - \Delta _{\theta <},\, \theta_{ms}
+ \Delta_{\theta >}) \times \, \nonumber \\ \Delta(\, \varphi,\, \varphi_{ms} -
\Delta_{\varphi <}, \, \varphi_{ms} + \Delta_{\varphi >}) \,
\label{windowSL}
\\
\mbox{with} \, \, \, \, \Delta (a,b,c) \,=\, S(a - b) \, - \, S(a
- c) \, , \nonumber
\end{eqnarray} 
where $f_{SL}$ is a constant (that is related to the ratio between the 
power entering the spillover and the
power entering the main beam and it is a number much less than
$1$ -- it will be estimated in Subsection \ref{comparisonamongcell}) and $S(x)$ is
the step function (or Heaviside function) that takes the value $1$
for $x \ge 0$ and the value $0$ otherwise. Equation~(\ref{windowSL}) is
just an asymmetrical rectangular box, in the $(\theta ,\varphi)$-plane,
centred around the point $(\theta_{ms},\varphi_{ms})$
and with sides of length $\Delta_{\theta >} + \Delta_{\theta <}$ and
$\Delta _{\varphi >} + \Delta _{\varphi <}$. 
Notice that the point
$(\theta_{ms},\varphi_{ms})$ is the direction of pointing of
the main spillover centre. 
%
%
%
This approximation leads to:
\be \int d \Omega \, B_{SL} = f_{SL}
\int_{\theta_{ms}-\Delta_{\theta <}}^{\theta_{ms}+\Delta_{\theta >}} 
\!\!\!\!\! d \theta
\sin
\theta \int _{\varphi_{ms}-\Delta_{\varphi <}}^{\varphi_{ms}
+\Delta_{\varphi >}}
\!\!\!\!\! d \varphi \, . \ee
Considering that
\begin{eqnarray} 
& & \int_{\theta_{ms}-\Delta_{\theta <}}^{\theta_{ms}+\Delta_{\theta >}}
\!\!\!\!\! d\theta \, \sin \theta \, \cos \theta = {1 \over 2} \sin \left( \delta + 2 \Delta_{\theta <}\right)
\times \nonumber \\  & &  \phantom{abcdefghilmno} \sin \left( \delta + 2 \,\theta_{ms}\right)\,
, \nonumber \\ 
& & \int_{\theta_{ms}-\Delta_{\theta <}}^{\theta_{ms}+\Delta_{\theta >}} 
\!\!\!\!\! d\theta \, \sin^2 \theta =
\Delta_{\theta <} + {\delta \over 2} -
\cos \left( \delta + 2 \, \theta_{ms} \right) \times \nonumber \\
& &  \phantom{abcdefghilmno} \, \sin \left( \delta + 2 \, \Delta_{\theta}\right)/2 \, , \nonumber \\
& & \int  _{\varphi_{ms}-\Delta_{\varphi <}}^{\varphi_{ms}
+\Delta_{\varphi >}} \!\!\!\!\!d \varphi \, \cos \varphi =
2 \cos \left( \varphi_{ms} \, + {\epsilon \over 2} \right) \times
\nonumber \\
& &  \phantom{abcdefghilmno} \sin \left( \Delta_{\varphi <} + {\epsilon \over 2}\right)\, , \nonumber \\ 
& & \int  _{\varphi_{ms}-\Delta_{\varphi <}}^{\varphi_{ms}
+\Delta_{\varphi >}} \!\!\!\!\! d \varphi \, \sin \varphi = 2 \sin \left( \varphi_{ms}+ {\epsilon \over 2} \right) \times \,
\nonumber \\
& &  \phantom{abcdefghilmno} \sin \left( \Delta_{\varphi <} + {\epsilon \over 2} \right) \, ,
\nonumber
\end{eqnarray}
with $\epsilon$ and $\delta$ implicitely defined by $\Delta_{\theta >} = \Delta_{\theta <} + \delta $
and $\Delta_{\varphi >} = \Delta_{\varphi <} + \epsilon $,
we obtain the final expression for the convolution:
\begin{eqnarray}
& & I^{TH} / f_{SL} = \nonumber \\  
& & {T_{10}\over 2} \sqrt{3 \over 4 \pi} \sin \left( \delta + 2 \Delta_{\theta}\right)
\sin \left( \delta + 2 \,\theta_{ms}\right)
(2 \Delta_{\varphi } + \epsilon ) -\nonumber \\
& & 4 \sqrt{3 \over 8 \pi} \left( \Delta_{\theta} + {\delta \over 2} -
\cos \left( \delta + 2 \, \theta_{ms} \right)
\, {\sin \left( \delta + 2 \, \Delta_{\theta}\right)\over 2} \right) \nonumber \\
& & Re \left[ T_{11} e^{i (\varphi_{ms}+\epsilon/2)}\right]
\sin \left( \Delta_{\varphi} + {\epsilon \over 2}\right)
\, , \label{convolution}
\end{eqnarray}
where the label $^{TH}$ stands for {\it Top Hat}.
Here we have made the notation lighter setting $\Delta_{\theta <} \equiv \Delta_{\theta }$ and $\Delta_{\varphi <} \equiv \Delta_{\varphi }$.
If the box is centred (i.e. $\delta =0$ and $\epsilon = 0$) and the direction of the main spillover
coincides with the spin axis [i.e. $(\theta_{ms},\varphi_{ms})=({\pi /2},
\varphi_{s})$], then the convolution is \citep{report2004}
\be {I^{TH} \over f_{SL}}= - \sqrt{6\over \pi} \left[ \Delta_{\theta} + 
{\sin 2\Delta \over 2} \right] \sin \Delta_{\varphi}
Re \left[ T_{11} e^{i \varphi_s(t)}\right]  \label{conv3}\ee
where we have made explicit the dependence on time for
$\varphi_s$. Notice that the term proportional to $T_{10}$ drops out
in this simple case.

\subsection{Gaussian approximation}
\label{gaussian}

The {\it Gaussian} approximation can be described rescaling
the dipole signal as
\be 
T_{1m} \rightarrow T_{1m} e^{-\sigma^2} \, ,
\label{dip_window}
\ee
where the factor $e^{-\sigma^2}$ is the Gaussian window function,
$e^{-\ell (\ell +1) \sigma^2/2}$, specified at $\ell=1$ and
$\sigma$ is 
the FWHM for the main spillover divided by $\sqrt{8 \, \ln 2 }$, 
and adopting as main spillover response 
\be 
B_{SL} (\theta, \varphi) = 
\, b \, \delta (\varphi , \varphi_{ms}){\delta (\theta , \theta_{ms}) \over \sin (\theta)} 
\, , \label{windowSL2}
\ee
where 
$b$ (as $f_{SL}$ before) is related
to the ratio between the power entering the main spillover 
and the power entering the main beam. 
In fact equation~(\ref{windowSL2}) is a {\it pencil beam} 
while the above rescaling of $T_{1m}$ mimics the
convolution of the dipole with a Gaussian beam.
In this way
it is easy to compute the following {\it Gaussian} convolution
\begin{eqnarray} 
& & I^{G} = 
\, b \, e^{-\sigma^2}\sqrt{{3\over {8 \pi}}} \times \nonumber \\
& & \left[ T_{10} \sqrt{2} \cos \theta_{ms} - 
2 Re\left[ T_{11} e^{i \varphi_{ms}}\right] \sin \theta_{ms}
\right]
\, . \label{gbeam}
\end{eqnarray}

\subsection{Link between top hat and Gaussian formalism}
\label{link}

We shall see in Subsection \ref{mainbeamandmainspillover} that we are interested in the 
following
directions for the mainspillover
\begin{eqnarray}
& & \theta_{ms} = {\pi \over 2} - \cos \theta _{mb} \, \alpha + {\cal O} \left( \alpha^3 \right)
\, , \label{thms}\\
& & \varphi_{ms} = \varphi \pm {\pi \over 2} \mp \sin \theta _{mb} \, \alpha + {\cal O} \left( \alpha^3 \right)
\, \label{phms} .
\end{eqnarray}
Replacing equations~(\ref{thms}) and (\ref{phms}) in equation~(\ref{convolution}) and considering a 
symmetrical {\it top hat} (i.e. $\epsilon = \delta = 0$ and 
$\Delta_{\varphi}=\Delta_{\theta}=\Delta $), one obtains the following expression
\begin{eqnarray}
& & I^{TH} = 4 \sqrt{{3\over 8\pi}} f_{SL} \left( \Delta + {\sin 2 \Delta \over 2}\right)
\sin \Delta \times \nonumber \\
& & \left\{ \pm Im\left[ T_{11}e^{i \varphi}\right] + 
{\Delta \sin (2 \Delta )\cos \theta \, \alpha \over { \sqrt{2} \left( \Delta + {\sin 2 \Delta \over 2}\right) \sin \Delta}} T_{10} \right. \nonumber \\
& & \left. - Re\left[ T_{11}e^{i \varphi}\right] \sin \theta \,\, \alpha 
\!\!\!\!\! \phantom{\sqrt{\Delta}\over{b (\Delta)} } \!\!\!\!\!\right\}
\, , \label{ITH} 
\end{eqnarray}
where only the linear term in $\alpha$ is kept.
Inserting equations~(\ref{thms}) and (\ref{phms}) in equation~(\ref{gbeam}) and 
expanding in $\alpha$ again to the linear order one finds
\begin{eqnarray}
& & I^{G} = 2 b \, e^{-\sigma^2}\sqrt{{3\over 8\pi}} \times \label{IG}
\\  
& & \!\!\!\!\!\!\!\! \left\{ \pm Im\left[ T_{11}e^{i \varphi}\right] + 
 {\cos \theta \, \alpha \over \sqrt{2}} T_{10} - Re\left[ T_{11}e^{i \varphi}\right] 
\sin \theta \, \alpha 
\phantom{a\over b}\!\!\!\! \right\}
\, . \nonumber 
\end{eqnarray}
Equation~(\ref{ITH}) and equation~(\ref{IG}) are very similar and become 
the same expression for small $\Delta$, if we define $b$ as follows
\be
b \, e^{-\sigma^2} = 4 f_{SL} \Delta^2 \, .
\ee
This is an intuitive result: up to normalizations, a {\it pencil beam}
and a {\it top hat beam} are not distinguishable when the (angular) area for
the top hat is sufficiently small. Given this result, in the following 
analytical treatment we will focus, for sake of compactness, 
only on the symmetrical {\it top hat} case adopting
\begin{eqnarray}
& & \epsilon = \delta =0 \, ,\\
& & \Delta_{\theta} = \Delta_{\varphi}= \Delta \, 
\end{eqnarray}
in the numerical estimates.
Instead,  we will widely consider the {\it Gaussian} case in
the section dedicated to the numerical simulations. 

\subsection{Relation between main beam and main spillover}
\label{mainbeamandmainspillover}

\begin{figure}

\centering

\includegraphics[width=6.7cm]{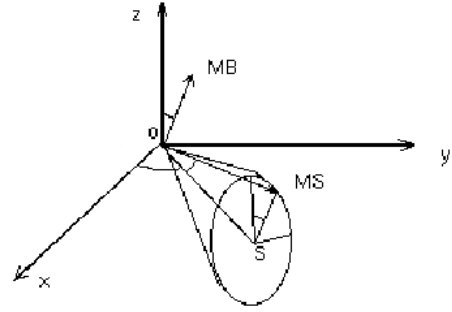}

\caption{Cone drawn by the direction of the main spillover (O-MS) during the rotation of the main beam (O-MB).
$\theta_{mb}$ is the angle between z axis and O-MB, $\varphi_s$ is the angle in x-y plane
between x axis and spin axis (O-S) and $\alpha$ is the angle between O-S and O-MS.}
\label{fig1}

\end{figure}

During the rotation of the main beam, the main spillover, if not lying 
on the spin axis,
draws a cone, as depicted in Figure~\ref{fig1}. This means that the 
direction of the main spillover and of the main
beam are related by
\begin{eqnarray}
& & \theta_{ms} = {\pi \over 2} - \arctan \left( \tan \alpha \, \cos \theta _{mb}\right) \, , \label{thetams}\\
& & \varphi_{ms} = \varphi_s \pm \arctan \left( \tan \alpha \, \sin \theta _{mb}\right) \, \label{varphims},
\end{eqnarray}
where $\theta _{mb}$ is the colatitude of the main beam, $\alpha$ is the 
angle at the vertex of the cone and $\varphi_s$ is the longitude of the 
spin axis (see also caption of Figure~\ref{fig1}).
The upper (lower) sign of equation (\ref{varphims}) has to be taken into account 
when the main beam rotates from North (South) to South (North). 

In order to make realistic this simple model, we have to check that during the rotation,
the solid angle, subtended by the main spillover ($\Omega_{ms}$), is constant.
A simple calculation gives
\be
\Omega_{ms} = 4 \Delta \sin \Delta \sin \theta_{ms}
= {4 \Delta \sin \Delta \over \sqrt{1+ \tan^2 \alpha \, \cos^2 \theta _{mb}}} \, .
\label{solidangle}
\ee
Of course equation~(\ref{solidangle}) is not constant because $\theta _{mb}$ 
is a function which depends on time.
However, we have 
\be
\Omega_{ms} = 4 \Delta \sin \Delta \left[ 1- {1 \over 2}\cos^2 \theta 
_{mb} \, \alpha^2 +
{\cal O}\left( \alpha^4 \right) \right]
\, ,
\ee
and therefore $\Omega_{ms}$ is constant in time to linear order in 
$\alpha$.
This means that the analytical calculation will be done to linear order 
in $\alpha$. Then
equations~(\ref{thetams}) and (\ref{varphims}) will be Taylor expanded 
for small $\alpha$, obtaining
\begin{eqnarray}
& & \theta_{ms} = {\pi \over 2} - \cos \theta _{mb} \, \alpha + {\cal O} \left( \alpha^3 \right)
\, , \label{thetamsexp}\\
& & \varphi_{ms} = \varphi_s \pm \sin \theta _{mb} \, \alpha + {\cal O} \left( \alpha^3 \right)
\, \label{varphimsexp}.
\end{eqnarray}

\subsection{Building the map}
\label{map}
\begin{figure}

\centering

\includegraphics[width=6.7cm]{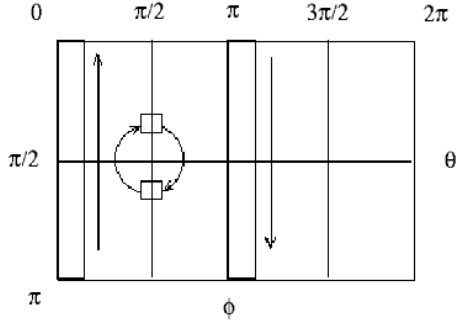}

\caption{Sketch of the nominal scanning during the first survey.}
\label{fig2}

\end{figure}

The total signal received by the satellite receiver, is the sum 
of the two following terms:
\be
T(\theta , \varphi) = T_{MB}(\theta ,\varphi) + I_{SL}(\theta,\varphi)
\, ;
\label{Ttot}
\ee
here $T_{MB}$ is the signal entering the main beam (where the
dipole has been subtracted away) whereas $I_{SL}$ is the signal
due to the dipole entering the main spillover.
We take into account for simplicity a simple scanning strategy as that considered
in the {\sc Planck} project as nominal scanning strategy (NSS) \citep{DupacTauber05},
i.e. with the spacecraft spin axis always in the antisolar direction 
or, in practice, always on the ecliptic plane as valid when the 
small effect of the spacecraft orbit around the Lagrangian
point L2 of the Sun-Earth system is neglected.

In this case,
for the first survey $I_{SL}$ 
is given by \be I^{(I)}_{SL}(\theta,\varphi)= \left\{
\begin{array}{c} I^{TH}\left( {\pi \over 2} - \alpha \cos \theta , \varphi + {\pi \over 2} -\alpha \sin \theta \right) \\ \;
\mbox{for} \; \; \; 0 < \varphi < \pi  \\  \\
I^{TH}\left({\pi \over 2} - \alpha \cos \theta , \varphi - {\pi
\over 2} +\alpha \sin \theta \right) 
\\ \; \mbox{for} \; \; \; \pi < \varphi < 2 \pi \, ,
\end{array}\right. \!\!\!\!\!\!\!\! \label{ISL1survey}
\ee
whereas, for the second survey it is
\be I^{(II)}_{SL}(\theta,\varphi)= \left\{
\begin{array}{c} I^{TH}\left( {\pi \over 2} - \alpha \cos \theta , \varphi - {\pi \over 2} + \alpha \sin \theta \right) \\ \;
\mbox{for} \; \; \; 0 < \varphi < \pi  \\  \\
I^{TH}\left({\pi \over 2} - \alpha \cos \theta , \varphi + 
{\pi \over 2} - \alpha \sin \theta \right) 
\\ \; \mbox{for} \; \; \; \pi < \varphi < 2 \pi \, .
\end{array}\right.  \!\!\!\!\!\!\!\!\label{ISL2survey}
\ee
The shift in the definition of the arguments in $I_{SL}$ (during either 
the first or the second survey) comes from the fact that
when the main beam rotates from North to South the main spillover
is shifted of $-\pi/2$ plus a small correction proportional to $ \alpha$
(due to the non-perfect alignment of the main spillover with the spin 
axis), while when the main beam rotates
from South to North the main spillover is shifted of $ +\pi/2$ minus
a small correction proportional to $\alpha$ (still due to the non-perfect alignment
of the main spillover with the spin axis).
Moreover note that it has been used that $\varphi_s = \varphi_{mb}-\pi/2$ when the main beam rotates from North to South while $\varphi_s = \varphi_{mb}+\pi/2$ when the main beam rotates from South to North.
See Figure~\ref{fig2} for a sketch of the scanning.

Notice that now $(\theta , \varphi)$ is the pointing of the main beam
(we omitted the label $mb$ to make the notation lighter).

\subsection{Definition of $T^{SL}_{\ell m}$}

As usual, we expand the signal in spherical harmonics
\be T(\theta, \varphi) = \sum _{\ell m} T_{\ell m} Y_{\ell m}(\theta , \varphi) \, , \ee
which implies \be T_{\ell m} = \int d\Omega \, T(\theta , \varphi) \,
Y_{\ell m}^{\star}(\theta , \varphi) \,. \label{coeff}\ee
We consider the multipole expansion of the straylight contribution 
\be
T^{SL}_{\ell m} = \int d\Omega \, I_{SL}(\theta, \varphi) \,
Y_{\ell m}^{\star}(\theta , \varphi)
\, . \label{TSL}
\ee
\noindent
To make lighter the notation, we rewrite the convolution in a compact form as
\be
& & I^{TH}(\theta_{ms},\varphi_{ms}) =
c_1 \sin (2 \theta_{ms}) - \nonumber \\
& & \left( c_2 - c_3 \cos (2 \theta_{ms}) \right)
\left( d_1 \cos \varphi_{ms} - d_2 \sin \varphi_{ms} \right)
\, ,
\ee
with
\begin{eqnarray}
& & c_1 = \sqrt{3 / 4 \pi} \, f_{SL} \Delta \, \sin (2 \Delta) \, T_{10} \\
& & c_2 = 4 \sqrt{3 / 8 \pi} \, f_{SL}\, \Delta \\
& & c_3 = 4 \sqrt{3 / 8 \pi}\, f_{SL} \, \sin (2 \Delta) /2 \\
& & d_1 = \sin \Delta \, Re \left[ T_{11}\right] \\
& & d_2 = \sin \Delta \, Im \left[ T_{11}\right] \, .
\end{eqnarray}


\subsection{General results after an odd number of surveys}
\label{resultsoddnumbersurvey}

After one or an odd number of surveys it is possible to show that 
all the $T_{\ell m}^{SL}$ with odd $m$ vanish when $\alpha=0$,
as follows from:
\begin{eqnarray}
& & T_{\ell m}^{0} = \int_{0}^{\pi} d \theta \sin \theta 
\left[ \int_{0}^{\pi} d \varphi I({\pi \over 2},\varphi + {\pi \over 2})
Y_{\ell m}^{\star}(\theta, \varphi) + \right. \nonumber \\
& & \left. \int_{\pi}^{2 \pi} d \varphi I({\pi \over 2},\varphi - {\pi \over 2})
Y_{\ell m}^{\star}(\theta, \varphi) \right]
\, ,
\label{intonesurvey}
\end{eqnarray} 
where the label $^0$ stands for $\alpha =0$ and where the label $^{SL}$ has been dropped for simplicity.
Changing variable of integration [$ \varphi = \varphi^{\prime}-\pi$ in the first integration and $ \varphi = \varphi^{\prime}+\pi$ in the second integration of equation~(\ref{intonesurvey})] and considering that
\be 
Y_{\ell m}^{\star}(\theta, \varphi \pm \pi) = (-1)^m Y_{\ell m}^{\star}(\theta, \varphi) 
\, ,
\ee
then equation~(\ref{intonesurvey}) reads
\be
\left[ 1 - (-1)^m \right] T_{\ell m}^0=0
\, .
\label{intonesurvey2}
\ee
Equation~(\ref{intonesurvey2}) is identically satisfied when $m$ is even and demands
$ T_{\ell m}^0=0 $ when $m$ is odd.

Moreover, still considering equation (\ref{intonesurvey}), 
the integration over $x = \cos \theta$ vanishes 
when $Y_{\ell m}(\arccos x,\varphi)$ is odd with respect to $x$. 
This happens when $\ell $ is odd {\it and} $m$ is even
\footnote{$Y_{\ell m}(\arccos x,\varphi)$ is odd with respect to $x$
also when $\ell $ is even {\it and} $m$ is odd.}, i.e. : 
\be
T_{\ell m}^0 = 0 \,\,\,\,\, \mbox{if $\ell$ is odd {\it and} $m$ is even}
\, .
\label{int1survey3}
\ee
Equations~(\ref{intonesurvey2}) and (\ref{int1survey3})
show that only 
$
T_{\ell m}^0  
$
with $\ell$ even {\it and} $m$ even,
can be non-vanishing. 
In particular $T_{\ell m}^0=0$ if $\ell$ is odd, a property
that will be widely used throughout the article.

Considering $\alpha$ small but not zero then is possible to write 
$T_{\ell m}^{SL}$ as follows
\be
T_{\ell m}^{SL} = T_{\ell m}^0 + \alpha \,\, T_{\ell m}^1 \, ,
\ee
where 
the label $^{1}$ stands for first order in $\alpha$.
It turns out that
\begin{eqnarray}
& & T_{\ell m}^1 = \int_{-1}^{1}dx 
\int_0^{2 \pi} d \varphi 
\left[2c_1 x  \phantom{a \over b}-  \right. \nonumber \\ 
& & \left. (c_2 + c_3)(d_1 \cos \varphi - d_2 \sin \varphi)\sqrt{1-x^2}\right]
\nonumber \\
& & Y_{\ell m}^{\star}(\arccos x, \varphi)
\, ,
\label{Tlm1storder}
\end{eqnarray}
where $x=\cos \theta$.
For symmetry reasons it is possible to show that only
\begin{eqnarray}
& & T_{\ell 0}^1 =  \int_{-1}^{1}dx 
\int_0^{2 \pi} d \varphi 
2c_1 x Y_{\ell 0}^{\star}(\arccos x, \varphi)
\, ,
\label{Tlm1storder2}
\end{eqnarray}
and
\begin{eqnarray}
& & \!\!\!\!\!\! T_{\ell \pm 1}^1 =  \int_{-1}^{1}dx 
\int_0^{2 \pi} d \varphi 
(c_2 + c_3)(d_1 \cos \varphi - d_2 \sin \varphi)
\nonumber \\
& & \sqrt{1-x^2}
Y_{\ell \pm 1}^{\star}(\arccos x, \varphi)
\, ,
\label{Tlm1storder3}
\end{eqnarray}
can be different from zero when $\ell$ is odd.
It turns out that both equations~(\ref{Tlm1storder2}) and (\ref{Tlm1storder3})
vanish except the case $\ell =1$.
We calculate
\be
T_{1 0}^1 =  2 \sqrt{ 4 \pi \over 3} c_1
\, ,
\ee
and
\be
T_{1 \pm 1}^1 = \sqrt{2 \pi \over 3} (c_2+c_3) \, \left( \pm d_1 + i d_2 \right)
\, .
\ee
This means that the unique term that turns on (linearly in $\alpha$) when $\alpha$ is taken different from zero (and small) is the dipole term.
All the other terms deviates from $\alpha = 0$ results at least for $\alpha^2$.

\subsection{Surviving systematic effect after an even number of surveys }
\label{survivingsystematic}

After two (or an even number of) surveys the average 
\be
\hat T_{\ell m} = (T^{(I)}_{\ell m} + T^{(II)}_{\ell m})/2
\, ,
\label{media}
\ee
where the labels $^{(I)}$ and $^{(II)}$ stand for first and second 
survey respectively,
is different from zero only for $\ell =1$ (to the linear order in $\alpha$). 
This term vanishes when $\alpha=0$. This means that the whole effect averages to $0$ when $\alpha=0$.

In order to show this statement, consider that with replacements similar to the ones performed in 
Subsection \ref{resultsoddnumbersurvey} one obtains
\be 
\hat T_{\ell m} = \alpha I_{\ell m}
\, ,
\ee
where $I_{\ell m}$ turns out to be the same integral given in equation~(\ref{Tlm1storder}) and studied in the Subsection \ref{resultsoddnumbersurvey}.
Notice that the average over two surveys is equivalent to one survey result.

This demonstrates the statement made at the beginning of this Subsection
\footnote{The fact that DSC has null average after $2$ surveys when $\alpha=0$ can be 
shown also integrating over the time interval $\Delta t$ corresponding to $2$ surveys.}. 

\subsection{The computation of $T^{SL}_{\ell m}$}

In Subsection \ref{resultsoddnumbersurvey} it has been shown that 
the calculation of $T_{\ell m}^{SL}$ to linear order in $\alpha$ is equivalent to 
the one with $\alpha=0$ (i.e. $T_{\ell m}^0$) if we exclude the 
case $\ell =1$. Moreover in order to have the possibility of obtaining a 
non-vanishing $T_{\ell m}^0$, we must constrain $\ell$ {\it and} $m$ to be 
even. 

\subsubsection{$T^{SL}_{0 0}$}

Specifying $\ell=0$ and $m=0$ we obtain
\be
T^{SL}_{0 0} = {4 \over \sqrt{\pi}} d_1 (c_2 + c_3) + {\cal O}\left( \alpha^2 \right)
\, .
\label{T00}
\ee
The monopole deviates from the order $\alpha =0$ as $\alpha^2$.


\subsubsection{$T^{SL}_{1 m}$}

Specifying $\ell=1$ and $m=0$ we obtain
\be
T^{SL}_{1 0} = 2 \sqrt{4 \pi \over 3}  c_1 \alpha
\, .
\label{T10}
\ee
For $\ell=1$ and $m=\pm 1$ we have
\be
T_{1 \pm 1}^{SL} = {1 \over 2}\sqrt{8 \pi \over 3} \left( \pm d_1 + i d_2 \right)
(c_2 + c_3) \alpha
\, .
\label{T1pm1}
\ee
Equations~(\ref{T10}) and (\ref{T1pm1}) show that only in the simplified case in which the spin axis is parallel to the direction of the main spillover (i.e. $\alpha = 0$), we have
that the DSC map does not show dipole contribution
\footnote{We will see in Subsection \ref{nss} that 
equations~(\ref{T10}) and (\ref{T1pm1}) can be generalized for 
non-small $\alpha$ substituting $\alpha$ with $\sin \alpha$.}.
%
%
%
This could have some important consequences on calibrations based on the kinematic dipole.
It is interesting to note that this contribution survives even after two surveys.
This is the unique contribution that does not average to zero (when $\alpha \neq 0$)
after an even number of surveys.

\subsubsection{$T^{SL}_{2 m}$}

Specifying $\ell=2$ and $m=0$, it is easy to obtain
\be
T^{SL}_{20} = {\cal O}\left( \alpha^2 \right)
\, ,
\label{T20}
\ee
implying that it is vanishing at linear order in $\alpha$.
Notice that even if symmetry arguments would make possible a contribution 
(i.e. $\ell$ and $m$ are both even) at the $0^{th}$ order in $\alpha$, the result is vanishing. 
As it is already known from Subsection \ref{resultsoddnumbersurvey} for $m= \pm 1$:
\be
T^{SL}_{2\pm1} = 0
\, ,
\label{Tpm1}
\ee
where no expansion in $\alpha$ has been performed in order to obtain this
result.
For $m=\pm2$:
\be
T^{SL}_{2 \pm 2} = - \left({4 \over 3}\right)^2 \sqrt{{15 \over 32 \pi}}
\left( d_1 \pm 2 i d_2 \right) \left( c_2 + c_3 \right)
\, ,
\label{T2pm2}
\ee
where $\alpha $ does not appear to linear order [there are corrections
of order ${\cal O}\left( \alpha^2 \right)$ ].

These results have been computed using the definition
of spherical harmonics for $\ell=2$, that we report here
for sake of completeness \citep{sakurai}:
\begin{eqnarray} 
& & Y_2^0 (\theta, \varphi) = \sqrt{5 \over {16 \pi }} (3 \cos^2 \theta -1)
\, , \\
& & Y_2^{\pm 1} (\theta, \varphi) = \mp \sqrt{15 \over {8 \pi }} e^{\pm i \varphi }
\sin \theta \cos \theta \, , \\
& & Y_2^{\pm 2} (\theta, \varphi) = \sqrt{15 \over {32 \pi }} e^{ \pm 2 i \varphi} \sin ^2 \theta
 \, .
\end{eqnarray}

Equation~(\ref{T2pm2}) is the (non-vanishing) contribution to the
quadrupole due to the dipole entering the straylight. This is one
of the main result of this work.

\subsubsection{$T^{SL}_{4 m}$}
Setting $\ell=4$, $m=0$ and $m=\pm 1$, we obtain:
\be
T^{SL}_{40} = T^{SL}_{4 \pm 1} = 0
\, .
\label{T40pm1}
\ee
Notice that even if symmetry arguments do not prevent a contribution, 
for $\ell = 4$ and $m =0$, nevertheless the computation gives $0$.
For $m=\pm 2$:
\begin{eqnarray}
T^{SL}_{4 \pm 2} = - \sqrt{{5 \over {2 \pi}}}
\left( d_1 \pm 2 i d_2 \right) {4 \over 15} \left( c_2 + c_3 \right) + 
{\cal O} \left( \alpha^2 \right)
\, .
\label{T4pm2}
\end{eqnarray}
For $m=\pm 3$:
\be
T^{SL}_{4 \pm 3} = 0
\, .
\label{T4pm3}
\ee
For $m= \pm 4$:
\be
T^{SL}_{4 \pm 4} = -{12 \over 225} \sqrt{35 \over {2 \pi}}
 \left( d_1 \pm 4 i d_2 \right) (c_2 + c_3)
+ {\cal O}\left( \alpha^2 \right)
\, .
\label{T4pm4}
\ee

These results have been computed using the definition for
of spherical harmonics for $\ell=4$, that we report here
for sake of completeness \citep{sakurai}:
\begin{eqnarray}
& & Y_4^0 (\theta, \varphi) = {3 \over 16 \sqrt{\pi}} ( 3 - 30 \cos^2 \theta + 35 \cos^4 \theta)
\, , \\
& & Y_4^{\pm 1} (\theta, \varphi) = \mp {3 \over 8} \sqrt{5 \over {\pi }} e^{\pm i \varphi } \times \nonumber \\
& & \phantom{abcdefghil} \cos \theta \sin \theta \left( -3 + 7 \cos^2 \theta  \right) \, , \\
& & Y_4^{\pm 2} (\theta, \varphi) = {3 \over 8} \sqrt{5 \over {2 \pi }} e^{ \pm 2 i \varphi} \times \nonumber \\
& & \phantom{abcdefghil} \sin ^2 \theta \left( -1 + 7 \cos^2 \theta \right) \, , \\
& & Y_4^{\pm 3} (\theta, \varphi) = \mp {3 \over 8} \sqrt{35 \over {\pi }} e^{\pm 3 i \varphi}
 \cos \theta \sin ^3 \theta \, , \\
& & Y_4^{\pm 4} (\theta, \varphi) = {3 \over 16} \sqrt{35 \over {2 \pi }} e^{\pm 4 i \varphi}
 \sin ^4 \theta \, .
\end{eqnarray}

%
%
%
%
%
%
%
%
%

\section{Impact on the Angular Power Spectrum}
\label{implicationforcosmology}
\setcounter{equation}{0}

The quadrupole presents a contamination 
that does not depend on $\alpha$ (to the order we are performing the computation).
In this Section we perform a detailed analysis of this effect.

\subsection{Comparison among $C^{SL}_{\ell}$ with low $\ell$}
\label{comparisonamongcell}

We consider
\be f_{SL} = {\int d \Omega B_{SL} \over {4 \, \Delta \sin \Delta }} ={ p \over {4 \, \Delta
 \sin \Delta }} \, ,
 \label{fsl}
 \ee
where $p$ is the
relative power entering the main spillover with respect to the total one
(i.e. essentially entering the main beam).
By the definition of $C^{SL}_{\ell}$,
\be C^{SL}_{\ell} = {1 \over {2 \ell + 1}}
\sum _{m=-\ell}^{\ell} \left( T_{\ell m}^{SL} \right)^{\star} T^{SL}_{\ell m}
\, ,
\ee
we compute
\be
C^{SL}_0 = 6 \left({p \over{\pi}}\right)^2 \, f(\Delta) Re \left[ T_{11} \right] ^2
\, ,
\ee
\be
C^{SL}_1 = {\alpha^2 p^2 \over{3}}
\left[ \cos^2 \Delta \, T_{10}^2 +
{1 \over 2}f(\Delta)
|T_{11}|^2
\right]
\, ,
\ee
\be
C^{SL}_2 = {2 \over{9}} \left({p\over \pi}\right)^2
f(\Delta)
\left( Re \left[ T_{11} \right]^2 + 4 Im \left[ T_{11} \right]^2\right)
\, ,
\ee
\be
C^{SL}_4 = \left({8 p \over{15 \pi}}\right)^2
{f(\Delta) \over 15}
\left( Re \left[ T_{11} \right]^2 + { 53\over 8} Im \left[ T_{11} \right]^2 \right)
\, ,
\ee
where
$$ f(\Delta)=\left( 1 + {\sin \Delta  \over \Delta} \cos \Delta  \right)^2 \, .$$
As an example, we choose $\Delta = \pi / 10$ and $p = 1/100$.
Moreover it is possible to show that 
\begin{eqnarray} 
& & Im \left[ T_{11} \right] = \sin \varphi_d \, \sin \theta_d \, \sqrt{2 \pi
\over 3} \, T \, , \\  
& & T_{10} = \cos \theta_d \, \sqrt{4 \pi
\over 3} \, T \, , \\ 
& & Re \left[ T_{11} \right] = - \cos \varphi_d \, \sin
\theta_d \, \sqrt{2 \pi \over 3} \, T \, , 
\end{eqnarray} 
where $(\theta_d,\varphi_d)$ is the direction and $T$ is the amplitude
of the dipole \footnote{These relations are obtained solving
the following set of equations
$T=T_{1 m} Y_{1}^{m}(\theta_d,\varphi_d)$,
$0=T_{1 m} Y_{1}^{m}(\theta_d + \pi/2,\varphi_d)$ and
$0=T_{1 m} Y_{1}^{m}(\pi /2,\varphi_d + \pi /2)$.
Of course, the solution can be verified replacing
in $T=T(\theta_d,\varphi_d)=T_{1 m} Y_{1}^{m}(\theta_d,\varphi_d)$.}.

In ecliptic coordinates, according to WMAP \citep{bennett}, 
$(\theta_d,\varphi_d)= (1.7651 \, , 2.9941) {\rm rad}$ and $T=3.346 \, {\rm mK}$.
We then obtain:
\begin{eqnarray} 
& & Im \left[ T_{11} \right] = 0.69823 \, {\rm mK} \, , \\  
& & T_{10} = -1.32225 \, {\rm mK} \, , \\ 
& & Re \left[ T_{11} \right] = 4.69963 \, {\rm mK} \, . 
\label{ReT11}
\end{eqnarray}

\noindent

Now we give, as an axample, some numerical results.
We have
\begin{eqnarray}
& & C^{SL}_0 = 5029.9 \, \mu{\rm K}^2 \\
& & C^{SL}_2 = 202.74 \, \mu{\rm K}^2 \\
& & C^{SL}_4 = 18.222 \, \mu{\rm K}^2 \, ,
\end{eqnarray}
and for $\alpha = \pi /36$
\be
C^{SL}_1 = 11.135 \, \mu{\rm K}^2
\, ,
\ee
while for $\alpha = \pi /18$
\be
C^{SL}_1 = 44.539 \, \mu{\rm K}^2
\, .
\ee
%


\subsection{The angular power spectrum}
\label{powerspectrum}
%
%

%


From equations~(\ref{Ttot}) and (\ref{coeff}) it is clear that: 
\be
T_{\ell m}=T_{\ell m}^{SKY}+ T_{\ell m}^{SL} 
\, , \label{sumofT} 
\ee 
where
$T_{\ell m}^{SL}$ is already defined in equation~(\ref{TSL}) and
$T_{\ell m}^{SKY}$ is 
\be 
T_{\ell m}^{SKY} = \int d\Omega \, T_{MB}(\theta ,
\varphi) \, Y_{\ell m}^{\star}(\theta , \varphi) 
\, .
\label{coeffsky}
\ee
The CMB power spectrum is given by \be C_{\ell} = {1 \over {2 \ell + 1}}
\sum _{m=-\ell}^{\ell} T_{\ell m}^{\star} T_{\ell m} \, , \ee and replacing
equation~(\ref{sumofT}) one obtains 
\begin{eqnarray}  
C_{\ell} &=& {1 \over {2 \ell + 1}} \sum _{m=-\ell}^{\ell} 
\left[ T_{\ell m}^{SKY} + T_{\ell m}^{SL} \right]^{\star} \left[T_{\ell m}^{SKY} + T_{\ell m}^{SL}\right] \nonumber
\\
&=& {1 \over {2 \ell + 1}} \sum _{m=-\ell}^{\ell} \left[ \left(T_{\ell m}^{SKY}
\right)^{\star} T_{\ell m}^{SKY} + \right.
\nonumber \\
& & \left. \left(T_{\ell m}^{SL}\right)^{ \star}
T_{\ell m}^{SL} + \left(T_{\ell m}^{SL}\right)^{\star} T_{\ell m}^{SKY} +
\left(T_{\ell m}^{SKY}\right)^{ \star}T_{\ell m}^{SL} \right] \nonumber
\\ \nonumber \\
&\equiv&  C_{\ell}^{SKY} + C_{\ell}^{SL} + C_{\ell}^{SKY-SL} \label{CMBps} \, .
\end{eqnarray}
In particular for the quadrupole we have 
\be
C_2^{SL} = {2 \over 5} F_{SL}^2 \left( \left[{Re\left[
T_{11}\right]}\right]^2 + 4 \, \left[{Im \left[ T_{11}\right]}\right]^2
\right)
\, ,
\label{c2sl_stat}
\ee
and
\be C_2^{SKY-SL} = - {4 \over 5} F_{SL} \left(Re \left[ T_{11}\right] Re
\left[ T_{22}^{SKY} \right] + \right. \nonumber \\ \left. 
2 Im \left[ T_{11} \right] Im \left[
T_{22}^{SKY} \right] \right)
\, ,
\label{C2SKYSL}
\ee
where it has been used 
$Re \left[ T_{2 \pm 2} \right] = Re \left[ T_{2 \mp 2} \right]$
and
$Im \left[ T_{2 \pm 2} \right] = - Im \left[ T_{2 \mp 2} \right]$,
with $F_{SL}$ defined as
\begin{eqnarray} 
F_{SL} &=&
f_{SL} {4 \over 3 \pi } \sqrt{5} \left[ \Delta
+ \cos \Delta \sin \Delta \right] \sin
\Delta \, \nonumber \\
&=& p \, {\sqrt{5}\over 3 \pi } 
\left[ 1 + \cos \Delta {\sin \Delta \over \Delta}\right]
\, .
\label{FSL}
\end{eqnarray}
From equation~(\ref{FSL}), choosing $\Delta = \pi / 10$ and considering $p = 1/100$ we have 
$ F_{SL}(p=1/100) \simeq 4.59 \times 10^{-3} $.
\subsection{DSC of the quadrupole}
\label{num_est}


We are ready to estimate the order of magnitude of $C^{SL}_2$ and
$C^{SKY-SL}_2$. Since $T^{SKY}_{2m}$ are stochastic numbers with vanishing mean and
standard deviation equal to $C_2^{SKY}$, we adopt, as an example, the following relation
\be
C_2^{SKY} \simeq 2 \, Re\left[ T_{22}^{SKY} \right]^2 = 2 \, Im\left[ T_{22}^{SKY} \right]^2
\,  \label{estimate}
\ee
for numerical estimate, where the factor 2 is due to the assumption that the real and imaginary part
give the same contribution
\footnote{This choice on the real and imaginary part of $T_{22}^{SKY}$ is limited to this Subsection. 
In Section \ref{statistic} a statistical analysis will be performed.}.
We obtain:
\be C^{SL}_2 = 202.7 \, \mu {\rm K} ^2 
\ee 
\be
C^{SKY-SL}_2 = \pm (386.1 \pm 114.7 ) \, \mu {\rm K} ^2 \label{CSKYSL2}
\ee 
where it has been 
chosen
$C_2^{SKY} \sim 10^3 \mu {\rm K}^2$.
The $\pm$ in equation~(\ref{CSKYSL2}) is due to our ignorance about the relative sign of $Re\left[
T_{22}^{SKY}\right]$ and $Im\left[T_{22}^{SKY}\right]$.

\begin{table*}
\begin{center}
\begin{tabular}{|c|c|c|c|c|}
\hline
$C_2(SKY)$ & $ p =1 / 500 $ & $ p = 1/100 $ & $p =5/ 100 $ & $C_2$ \\
\hline
$ 500 $ & $8.1$ & $202.7$ & $5068.5$ &  SL\\
\hline
$ 500 $ & $54.6 \pm 16.2$ & $273.0 \pm 81.1$ & $1364.9 \pm 405.6$ & SKY-SL\\
\hline
$ 1000 $ & $8.1$ & $202.7$ & $5068.5$ & SL \\
\hline
$ 1000 $ & $77.2 \pm 22.9$ & $386.1 \pm 114.7$ & $1930.3 \pm 573.6$ & SKY-SL\\
\hline
$ 1500 $ & $8.1$ & $202.7$ & $5068.5$ & SL\\
\hline
$ 1500 $ & $94.6 \pm 28.1$ & $472.8 \pm 140.5$ & $2364.1 \pm 702.5$ & SKY-SL\\
\hline
\end{tabular}
\end{center}
\caption{All the temperatures are given in $\mu {\rm K}^2$. See also the text.}
\label{one}
\end{table*}

In Table \ref{one} there are $C^{SL}_2$ and $C^{SKY-SL}_2$ for
$p=1/500 \, , 1/100, 5/ 100$ and $C_2^{SKY} = 500, \, 1000, \,
1500 \, \mu  {\rm K}^2$. All the numbers for $C_2^{SKY-SL}$ contribution have to be understood
with a $\pm$ in front of them (since we do not know the total sign of this contribution).

We stress two main observations.
First, we notice that the $C^{SL}_2$ contribution is smaller than the 
$C^{SKY-SL}_2$ contribution
if $p$ is sufficiently small. This is clear because $C^{SL}_2$ is 
quadratic in $p$ while $C^{SKY-SL}_2$ is linear.
Therefore, for $p=1/500$ and $p=1/100$ we have that the leading term is given by $C^{SKY-SL}_2$ while
for $p=5/100$ we find that $C^{SKY-SL}_2$ is subleading;
second, since $C^{SL}_2$ contribution is always positive, because of the 
previous observation, it is clear that $C_2^{SKY}$ can be lowered only for the first and second column of Table \ref{one}, when the relation (\ref{estimate}) is chosen.

\subsection{Decrease of the quadrupole}

From equation~(\ref{CMBps}) we can rewrite the observed quadrupole $C_2$ as a function of $F_{SL}$
\be
C_2 (F_{SL})= C_2^{SKY} - {4 \over 5} B \, F_{SL} + {2 \over 5} A \, F_{SL}^2
\, ,
\label{C2FSL}
\ee
where
\begin{eqnarray}
B \equiv Re \left[ T_{11}\right] Re
\left[ T_{22}^{SKY} \right] + 2 Im \left[ T_{11} \right] Im \left[
T_{22}^{SKY} \right] \\
A \equiv \left[{Re\left[
T_{11}\right]}\right]^2 + 4 \, \left[{Im \left[ T_{11}\right]}\right]^2 \, .
\end{eqnarray}
Equation~(\ref{C2FSL}) is just a parabolic behaviour in $F_{SL}$ with concavity in the upward direction
since the coefficient of $F_{SL}^2$ is always positive.
The sign of $B$ is not known a priori: from equation~(\ref{C2SKYSL}) or 
equation~(\ref{C2FSL}) is clear
that $C_2^{SKY}$ can decrease only if $B>0$.
We focus now on this case. 

\begin{figure}

\centering

\includegraphics[width=6.7cm]{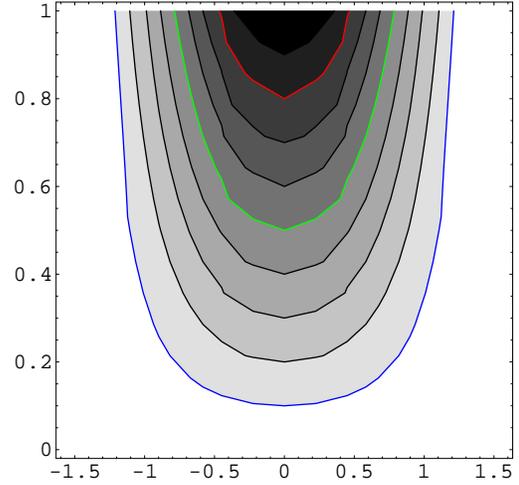}
\caption{Contour plot for the ratio $ C_2 (F_{SL}|_v) / C_2^{SKY}$ versus the couple of parameters 
$(\gamma ,f)$. Since we are dealing with $B>0$ than 
$\gamma \in [-\pi/2,\pi/2]$.
Each contour represents the value $n/10$ with $n \in [1,2,...,9]$.
The red contour represent $1/5$, the green contour $1/2$ and the blu contour $9/10$.}
\label{fig3}

\end{figure}

The minimum value $C_2 (F_{SL}|_v)$ is reached by the vertex $F_{SL}|_v$,
$ F_{SL}|_v = B/A
$, where
\begin{eqnarray}
& & \!\!\!\!\!\! C_2 (F_{SL}|_v) - C_2^{SKY} = - {2 \over 5} {B^2 \over A} \label{C2FSLV} \\
& & \!\!\!\!\!\! = -{2 \over 5} { \left( Re \left[ T_{11}\right] Re
\left[ T_{22}^{SKY} \right] + 2 Im \left[ T_{11} \right] Im \left[
T_{22}^{SKY} \right]\right)^2 \over \left[{Re\left[
T_{11}\right]}\right]^2 + 4 \, \left[{Im \left[ T_{11}\right]}\right]^2}
\, . \nonumber
\end{eqnarray}
Considering that
$
|T_{22}^{SKY}|^2 \in [0,5 C_2^{SKY}/2]
$, we can rewrite equation (\ref{C2FSLV}) as
\be
C_2 (F_{SL}|_v)=C_2^{SKY} \left[1-f \, \cos^2 \gamma \right]
\, ,
\ee
where $f = |T_{22}^{SKY}|^2/(5 C_2^{SKY}/2)$ (then $f \in [0,1]$) 
and $\cos \gamma ={(\vec v \cdot \vec u)/ v}$ with $v$ being the norm 
of the vector $\vec v$, 
\be
\vec v = \left(Re \left[ T_{11}\right],2 \, Im \left[ T_{11}\right]\right) \, ,
\ee
and $ \vec u = \left(\cos \alpha, \sin \alpha \right) $
with $\alpha$ defined as 
\be
\alpha = \arctan \left(Im\left[ T_{22}^{SKY} \right]/Re\left[ T_{22}^{SKY} \right] \right) \, ,
\ee
and therefore ranging in the set $[0,2 \pi]$.
In Figure \ref{fig3} we show a contour plot for the ratio
$ C_2 (F_{SL}|_v) / C_2^{SKY}$ versus the couple of parameters $(\gamma , f )$. Darker is the grey 
and smaller is the ratio (see also the caption).

\begin{figure}

\centering

\includegraphics[width=6.7cm]{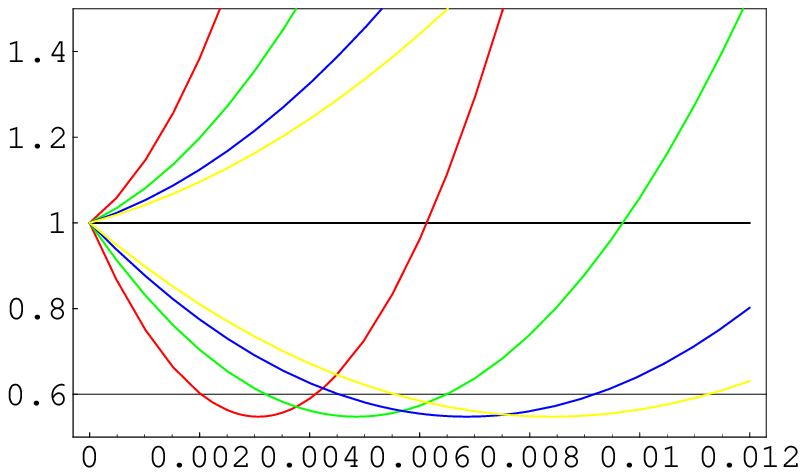}
\includegraphics[width=6.7cm]{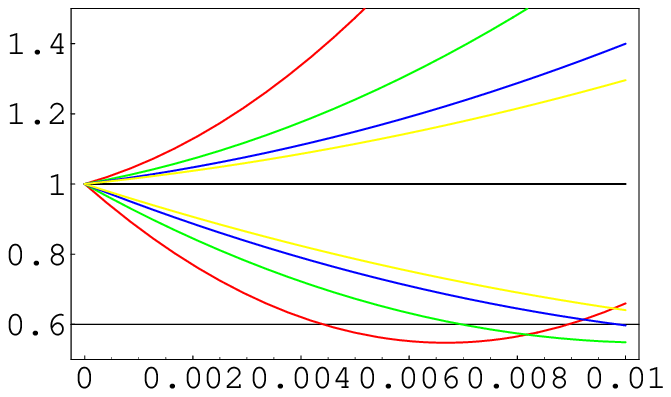}

\caption{Top panel: $y=y(F_{SL})$ for $C_2^{SKY} = 200 \, \mu {\rm K}^2$ (red line), $C_2^{SKY} = 
500 \, \mu {\rm K}^2$ (green line),
$C_2^{SKY} = 1000 \, \mu {\rm K}^2$ (blu line) and $C_2^{SKY} = 1500 \, \mu {\rm K}^2$ (yellow 
line). 
Lower curves correspond to $(\gamma,f) = (\pi/10,1/2)$ while upper ones to $(\gamma,f) 
=(\pi/10+\pi/2,1/2)$.
Bottom panel is a zoom of top panel but plotted in terms of $p$ instead of $F_{SL}$ (see equation 
(\ref{FSL})):
$y=y(p)$ for $C_2^{SKY} = 200 \, \mu {\rm K}^2$ (red line), $C_2^{SKY} = 500 \, \mu {\rm K}^2$ 
(green line), 
$C_2^{SKY} = 1000 \, \mu {\rm K}^2$ (blu line) and $C_2^{SKY} = 1500 \, \mu {\rm K}^2$ (yellow 
line),
where $\Delta = \pi /10$. Lower curves correspond to $(\gamma,f) = (\pi/10,1/2)$ 
while upper ones to $(\gamma,f) =(\pi/10+\pi/2,1/2)$. See also the text.}

\label{fig4}

\end{figure}

In the same way we rewrite equation (\ref{C2FSL}) as
\begin{eqnarray}
& & y(F_{SL}) = {C_2 (F_{SL})\over C_2^{SKY}} \label{ydiFSL} \\
&=& 1 -  f^{1/2} {v \, cos \gamma \over \sqrt{5 \, C_2^{SKY}}} \, F_{SL} 
+ {2 \over 5} {v^2 \over C_2^{SKY}} \, F_{SL}^2
\, .
\end{eqnarray}
This means that the first branch of DSC is given by 
$\gamma \in [-\pi/2,\pi/2]$ (where also a decreasing of the observed quadrupole is possible, i.e. 
$B>0$) and the second branch is given by $\gamma \in [\pi/2,3 \pi /2]$ (where only an increasing 
of the observed quadrupole is possible, i.e. $B<0$).
In Figure \ref{fig4} we plot equation (\ref{ydiFSL}) 
for $C_2^{SKY} = 200, 500, 1000, 1500 \, \mu {\rm K}^2$ (red, green, blu and yellow lines 
respectively) 
with $(\gamma,f) = (\pi/10,1/2)$ (lower curves) 
and with $(\gamma,f) = (\pi/10+\pi/2,1/2)$ (upper curves).
Bottom panel of Figure \ref{fig4} is a zoom of the top panel 
but with $p$ as independent variable
instead of $F_{SL}$ (see also the caption).

\section{Numerical Simulations}
\label{numericalsimulations}
\setcounter{equation}{0}

The results presented in the previous sections 
are based on an analytical treatment of the DSC effect.
Numerical simulations allow to independently 
check the analytical results of the previous Section and to extend 
the analysis to more complex (and realistic) cases 
by relaxing some of the simple assumptions adopted in the previous sections.
In particular, 
we will investigate 
through dedicated computations
some aspects that are difficult or impossible to address in analytical way:
i) the case of a non-small angle $\alpha$ 
(i.e. of a deviation of some tens of degrees 
from the orthogonality between the directions of the main beam and the 
main spillover); ii) more complicate scanning strategies
including, for instance, precessions of the spacecraft spin axis
around the nominal case of a spin axis direction kept along the antisolar
direction, i.e. in practice on the ecliptic plane.
Moreover, simulations allow to take into account iii) the effect introduced by the 
uncompleteness of one of the two surveys
and iv) to extend the analysis to higher multipoles
in the cases i) and ii).

\begin{figure*}
   \begin{tabular}{cc}
\includegraphics[width=5cm,angle=90]{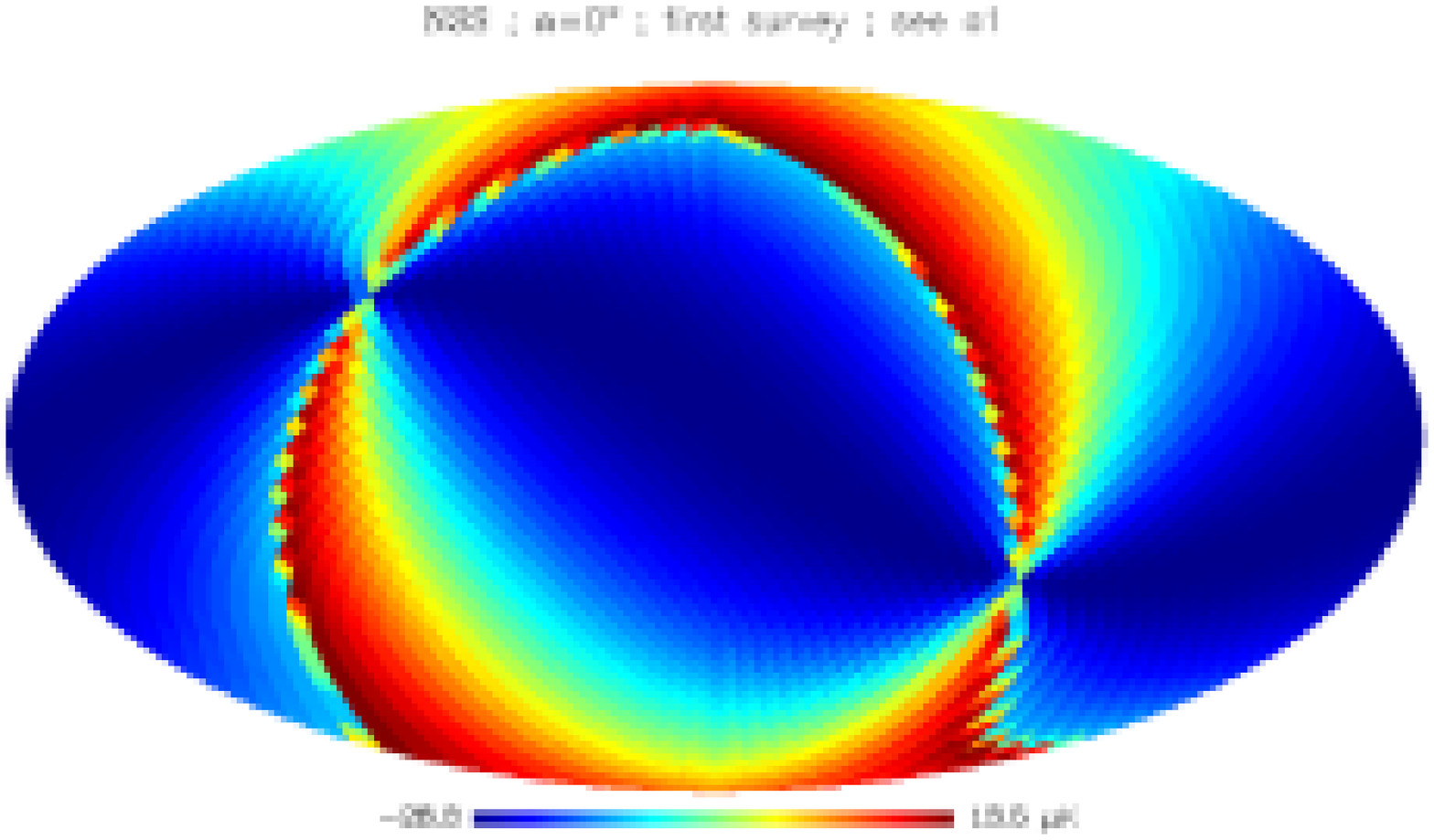}&
\includegraphics[width=5cm,angle=90]{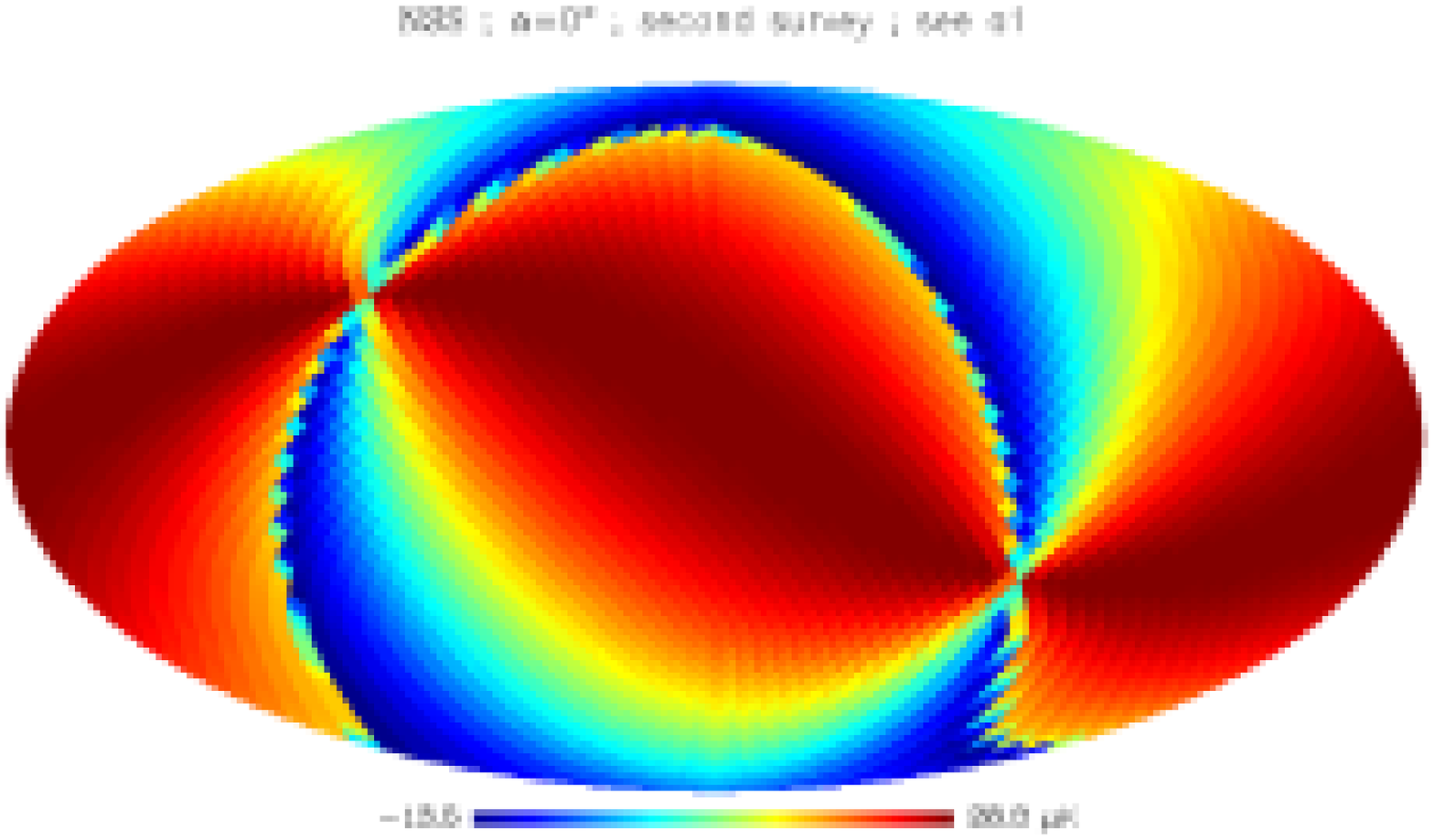}\\
\includegraphics[width=5cm,angle=90]{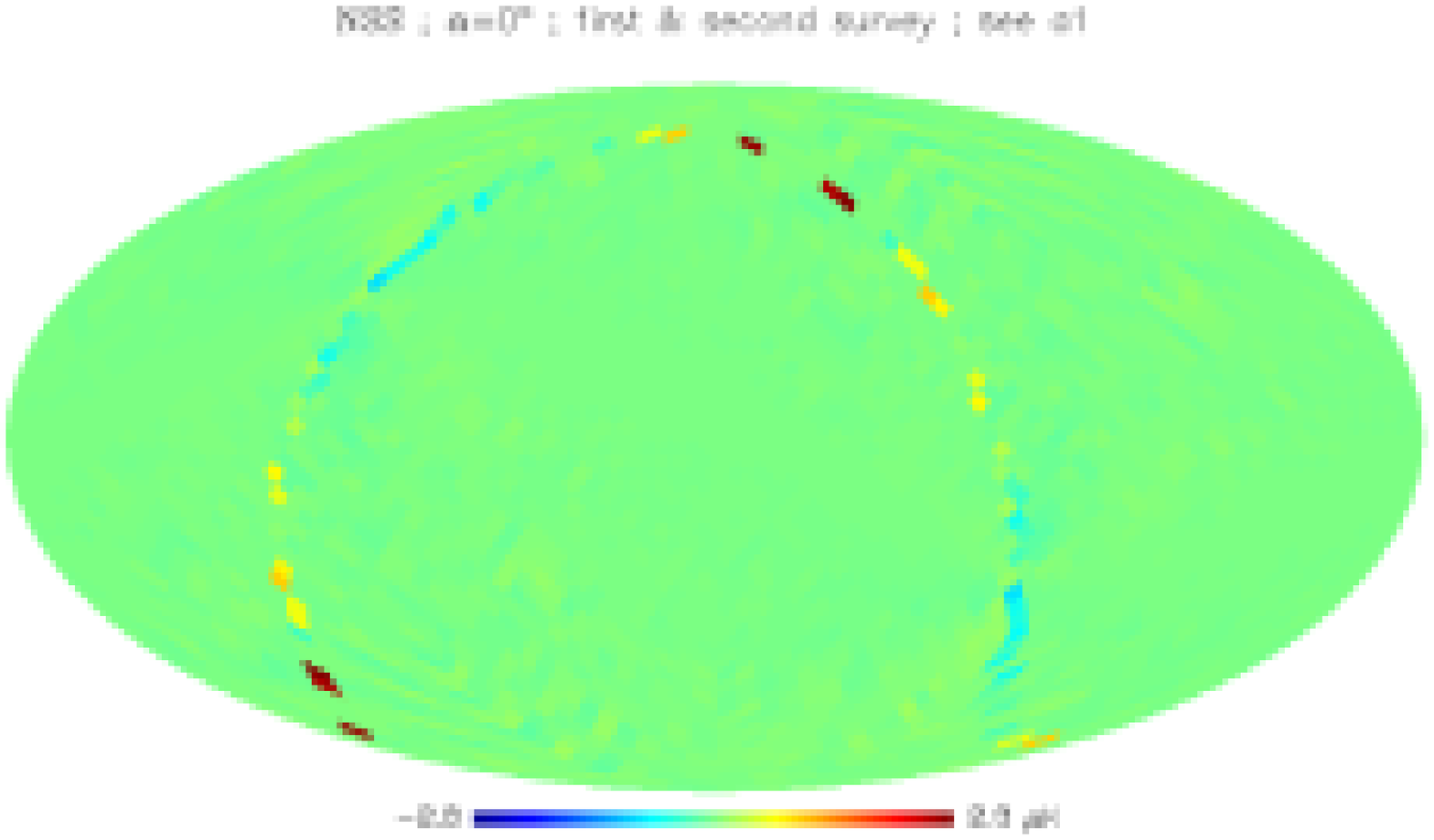}&
\includegraphics[width=5cm,angle=90]{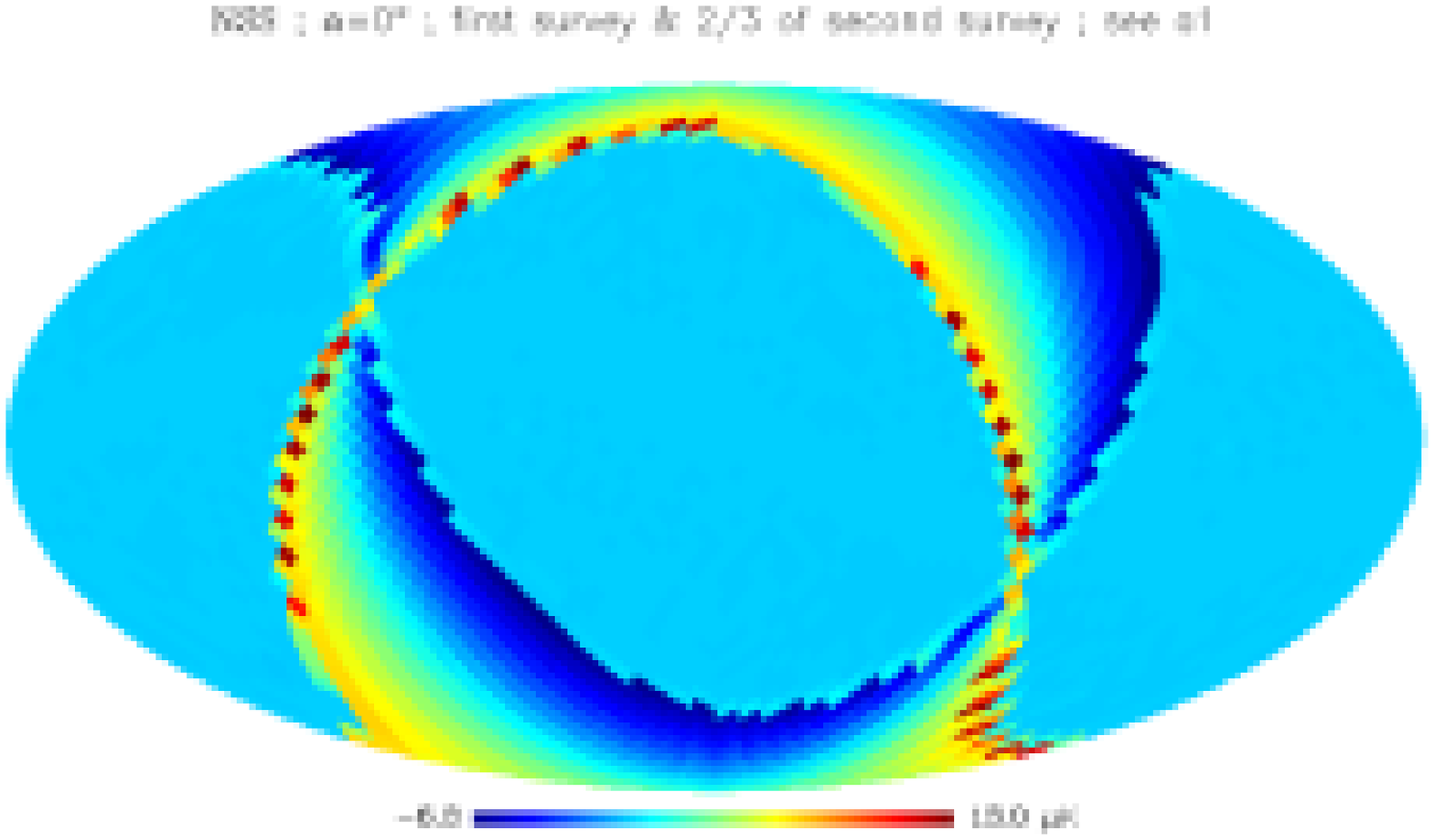}
   \end{tabular}
\caption{Maps of DSC in the case of the nominal scanning strategy (NSS)
and for $\alpha = 0^\circ$.
The parameters of each case are indicated 
above each map together with the reference to the panel in Figure~\ref{cl_allcases} where
the corresponding APS is displayed. 
Galactic coordinates and Mollweide projection are used.} 
\label{map_nss}
\end{figure*}

\begin{figure*}
   \begin{tabular}{cc}
\includegraphics[width=5cm,angle=90]{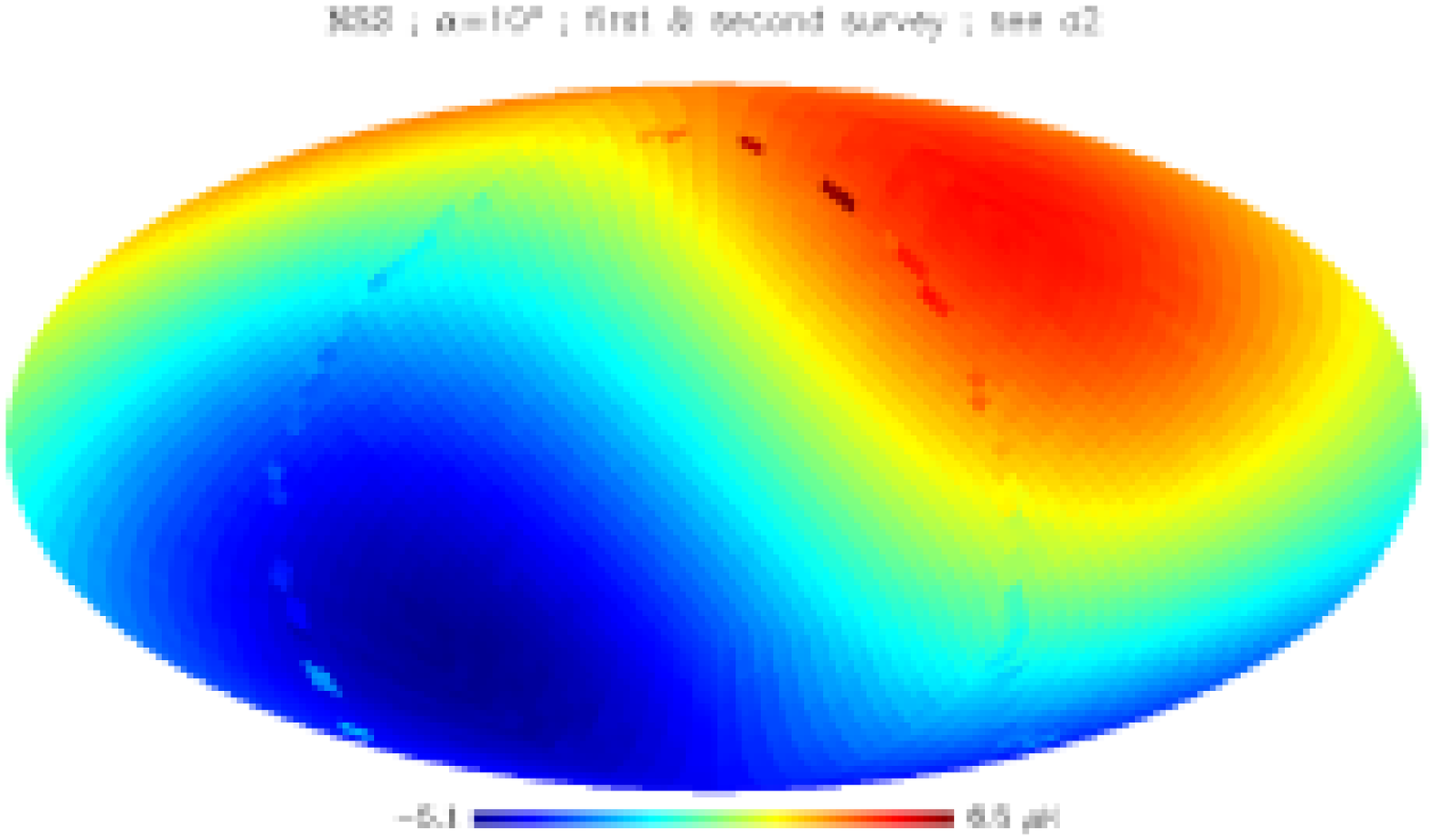}&
\includegraphics[width=5cm,angle=90]{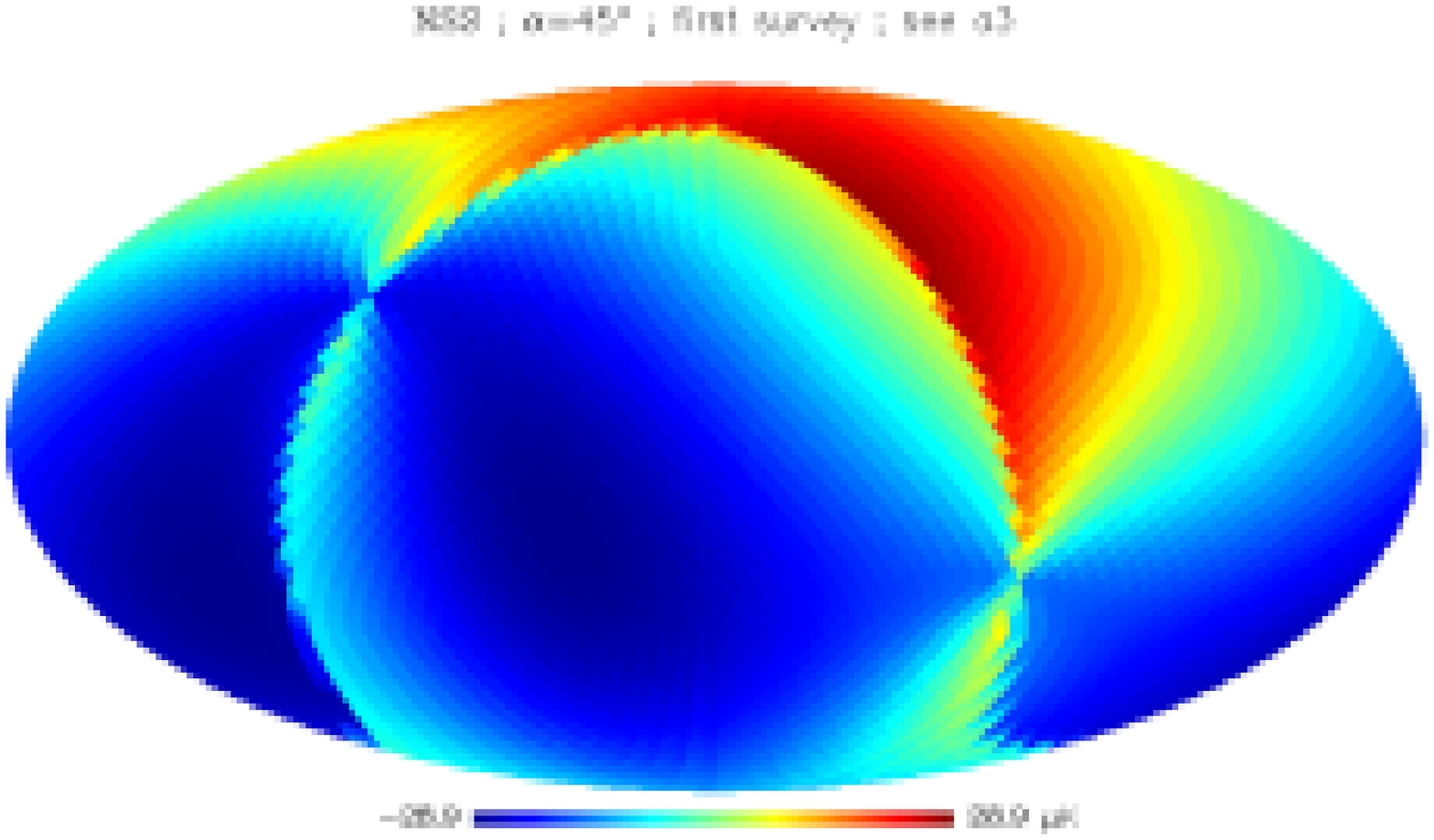}\\
\includegraphics[width=5cm,angle=90]{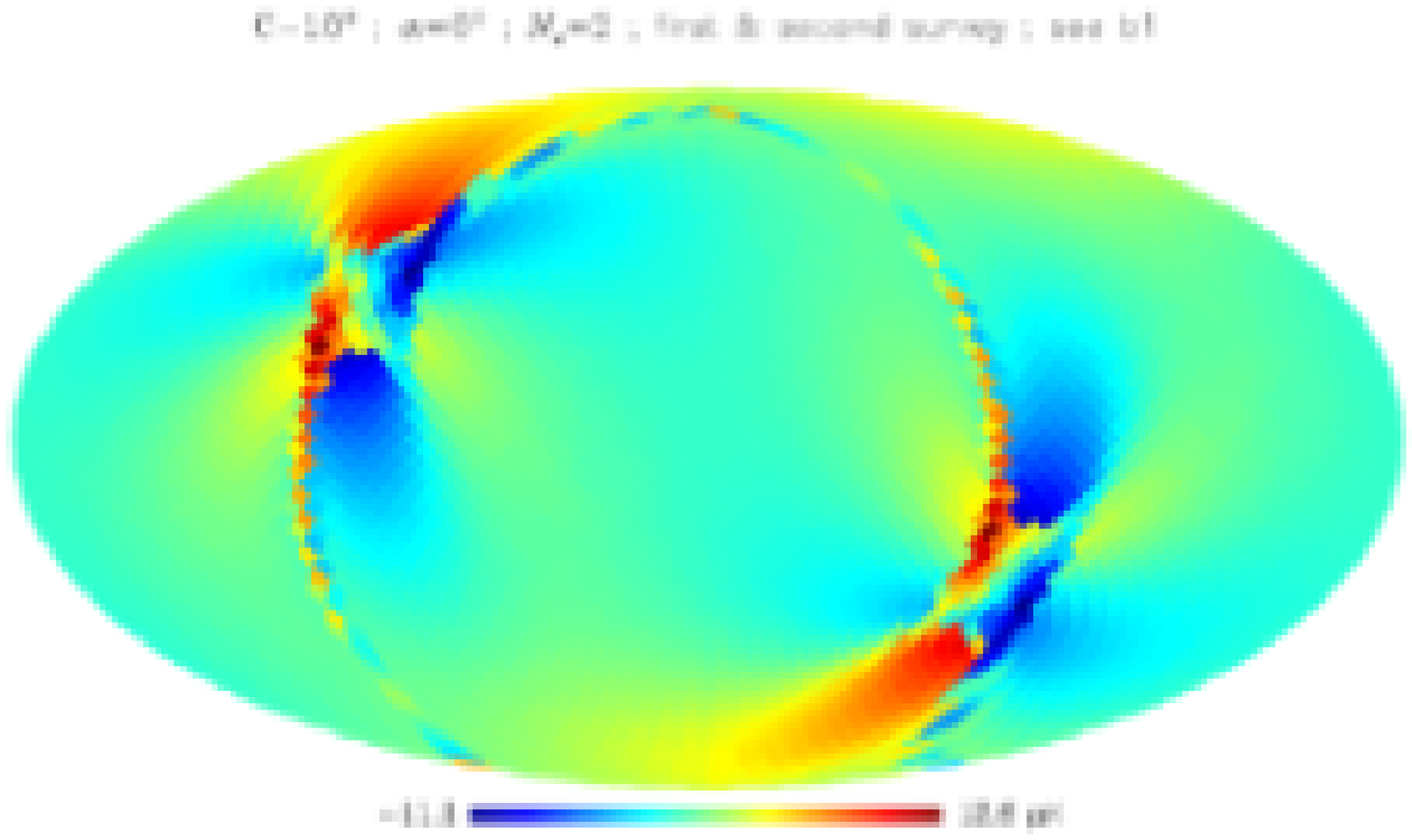}&
\includegraphics[width=5cm,angle=90]{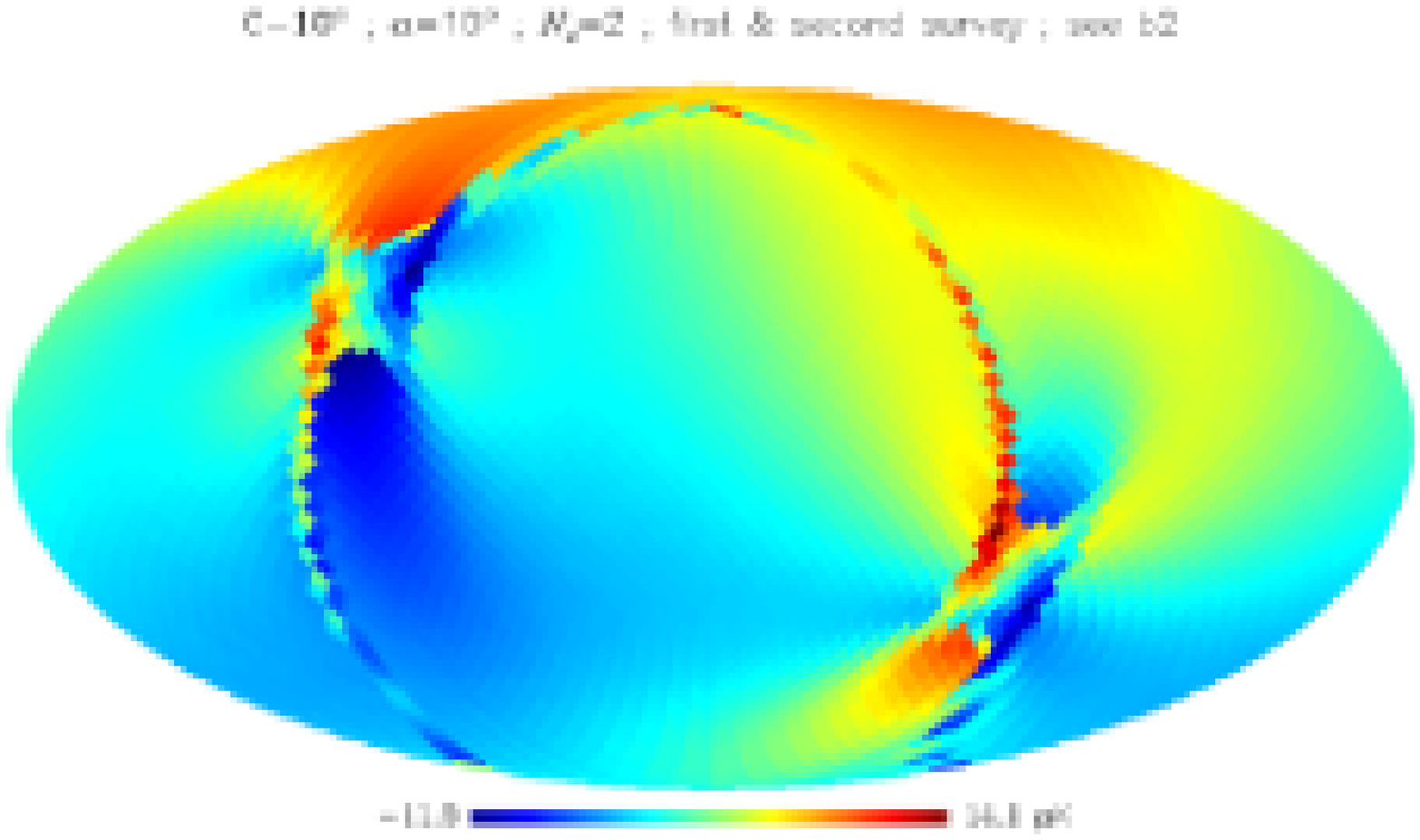}
   \end{tabular}
\caption{Maps of DSC in the case of the nominal scanning strategy (NSS)
and for $\alpha \ne 0^\circ$ and in the case of a cycloidal scanning strategy (C) 
with a semi-amplitude of $10^\circ$ and $\alpha = 0^\circ$ or $10^\circ$.
The parameters of each case are indicated 
above each map together with the reference to the panel in Figure~\ref{cl_allcases} where
the corresponding APS is displayed.
Galactic coordinates and Mollweide projection are used.} 
\label{map_nss_c}
\end{figure*}

\subsection{The numerical code}
\label{numericalcode}

We used here an update
dedicated version of the code implemented 
and successfully tested in many simulation works 
devoted to the  study (and reduction) of various classes of
systematic effects in the context of the {\sc Planck} mission
under different scanning strategy assumptions. 
It is described in detail in \citet{burigana98} and \citet{maino99} 
and, in particular regarding the straylight effect, in \citet{burigana1,burigana2}, 
where further informations on the relevant reference systems adopted in the code
can be found.
We do not consider here the impact of the spacecraft orbit (around the Lagrangian
point L2 of the Sun-Earth system, as for example in the case of WMAP and {\sc Planck})
because its effects are not relevant in this context.

In this work, we compute the convolutions between the main spillover response
and the sky dipole signal as described in \citet{burigana1,burigana2},
but by pixellizing the sky at $\simeq 1^\circ$ 
resolution~\footnote{The {\tt HEALPix} scheme (http://healpix.jpl.nasa.gov/) 
by \citet{gorski05} has been adopted
in simulations. A dipole input map at $n_{\rm side} =64$ has been considered.},
considering spin-axis shifts of $\sim 1^{\circ}$ every day
and 396 samplings per scan circle, and by adopting the analytical Gaussian
description of the main spillover response introduced 
in Section~\ref{gaussian}~\footnote{This is certainly a simple approximation of the
complexity of the realistic shapes predicted for the main spillover 
(and in general for the beam response far from the main beam)
by optical simulation codes (see e.g. \citet{sandri}).
On the other hand, the details of the main spillover response shape
depend on the considered optical system. 
In the context of the {\sc Planck} activities, 
they will be included in a future work.
The main contribution of this paper is in fact the understanding
of the most relevant effects introduced by the DSC, 
common to relatively different optical systems,
in terms of simple parametrizations.}
that has been explicitely implemented in this version of the code.

We consider here as reference case a channel at 70~GHz, a frequency 
where the foreground contamination at large angular scales 
is minimum \citep{bennettfore,page}, 
and report simulation results in terms of antenna temperature, as typical
in simulation activity being it an additive quantity with respect to the sum
of various contributions. 
For numerical estimates, we report here the results referring to the case 
$p=0.01$ (i.e. a relative power of 1\% entering the main spillover)
and assume a main spillover FWHM, FWHM$_{ms}$, of $20^\circ$, an angular size 
amplitude comparable to those of CMB space experiments \citep{sandri,barnes}.

The adopted resolution allows us to investigate
the DSC effect on angular scales larger than few degrees,
i.e. up to multipoles $\ell \sim 50$, where the most relevant effects are 
expected because of the large scale features considered here 
both for the signal and the main spillover response.
The main output of the simulation code consists of time ordered data
(TOD) containing the signal entering the main spillover for each simulated pointing
direction. We generate TOD for two complete sky surveys.
The TOD are then 
coadded to produce all sky maps of DSC
at a resolution of $\simeq 2^\circ$ ($n_{\rm side} =32$ in 
the {\tt HEALPix} scheme). These maps are analysed in terms of 
angular power spectrum (APS) by using the {\tt anafast} facility
of the {\tt HEALPix} libraries.
For simplicity, we considered always an angle of $90^\circ$
between the directions of the spacecraft spin axis and of the main beam centre
(assumed to be aligned with the telescope line of sight).
This assumption allows to reach the all sky coverage for the whole set of 
considered scanning strategies. Differently,
in the absence of spin axis displacements from the ecliptic plane 
a small unobserved circular patch appears around each of the ecliptic poles
\citep{DupacTauber05}. We prefer to avoid the case
of non-complete sky coverage because it may generate a certain complication 
in the APS analysis,
introducing a mix between the effects from 
DSC and partial sky coverage.
Of course, this choice allows also to check the analytical 
results in the same work condition.

Our general results are obviously not significantly dependent
on the specific choice for FWHM$_{ms}$. In addition, 
they can be easily rescaled 
to any value of $p$ (signal $\propto p$, APS $\propto p^2$).

Finally, we note that the discretization of $\simeq 1^\circ$ 
adopted in the present simulations implies
a typical numerical error amplitude significantly less (because of averaging) than
$\sim 3~{\rm mK}~\times~(0.5 ^\circ / 180^\circ)~\times~p~\sim 0.1~\mu$K,
a value at least one or two order of magnitude smaller than those 
relevant in this analysis, and then fully negligible for the present 
purposes.
 
\begin{figure*}
   \begin{tabular}{cc}
\includegraphics[width=5cm,angle=90]{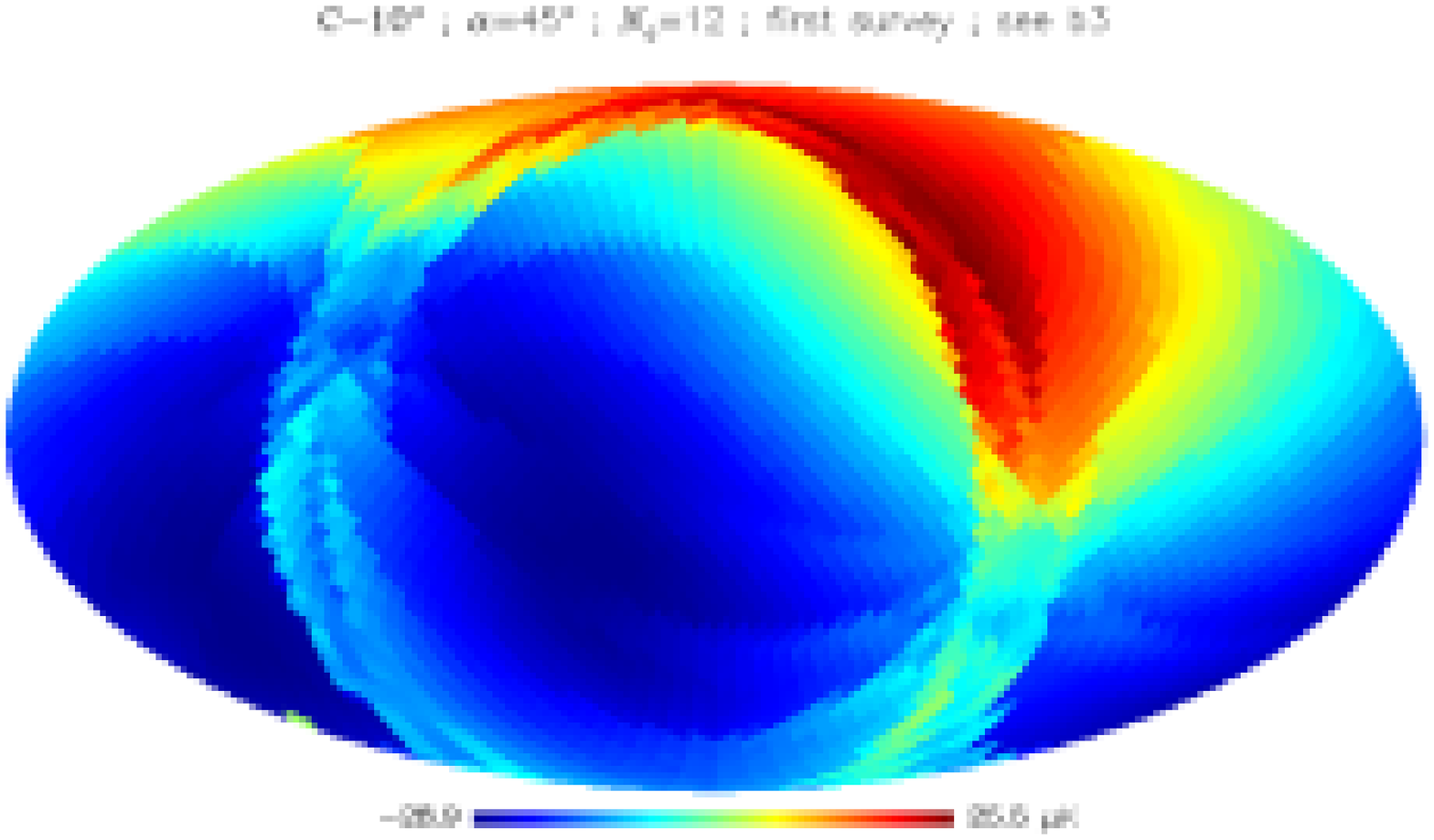}&
\includegraphics[width=5cm,angle=90]{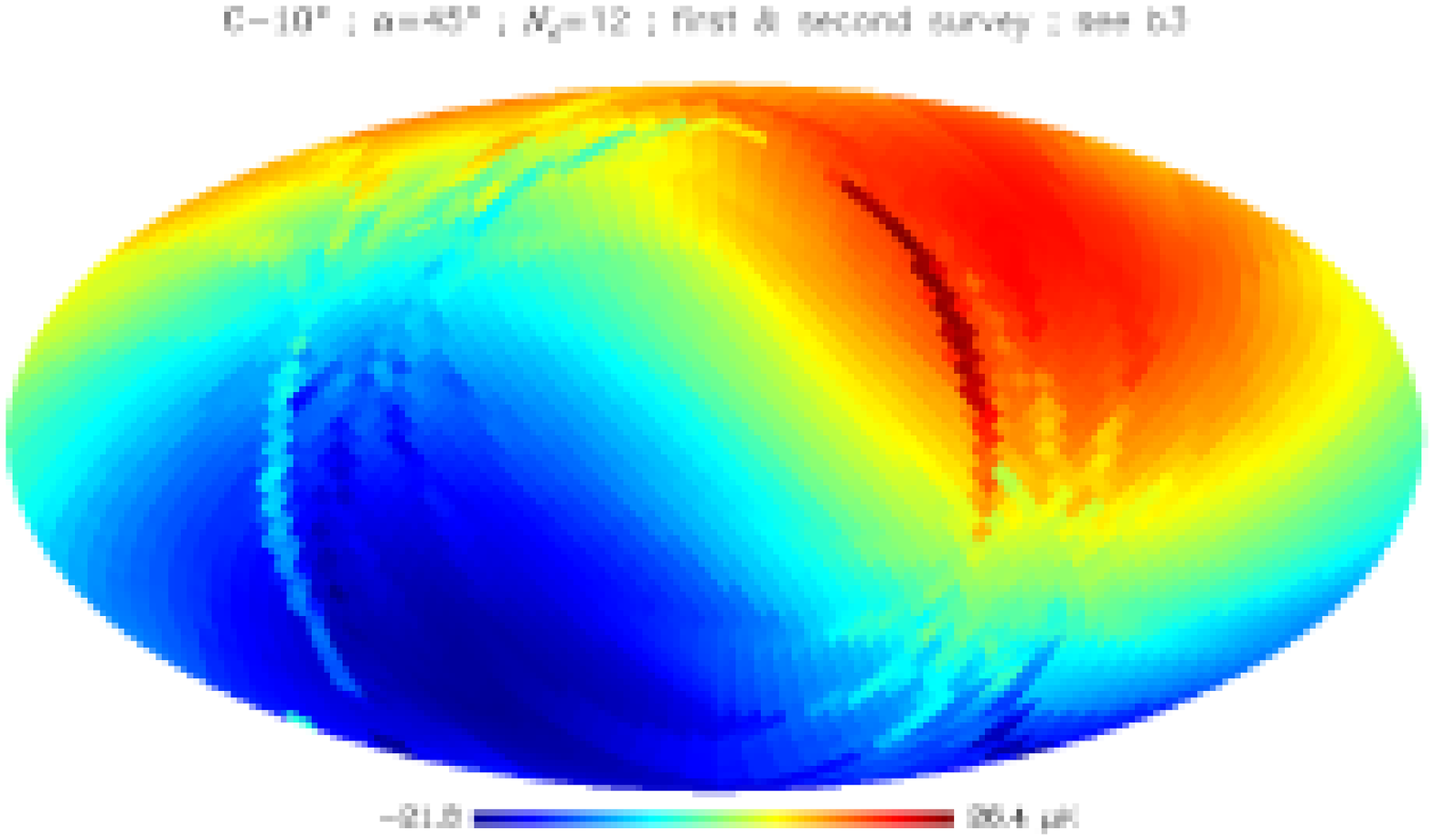}\\
\includegraphics[width=5cm,angle=90]{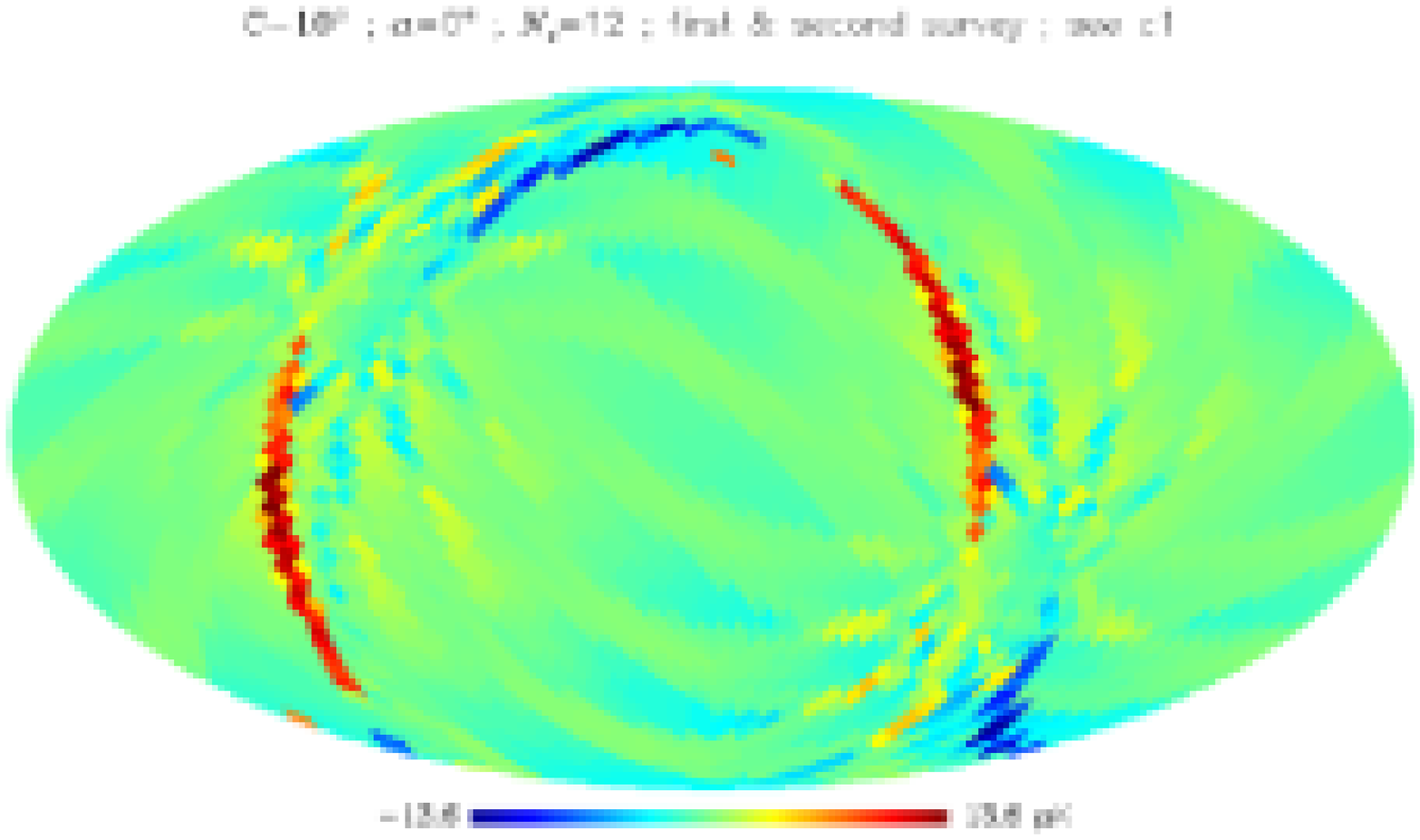}&
\includegraphics[width=5cm,angle=90]{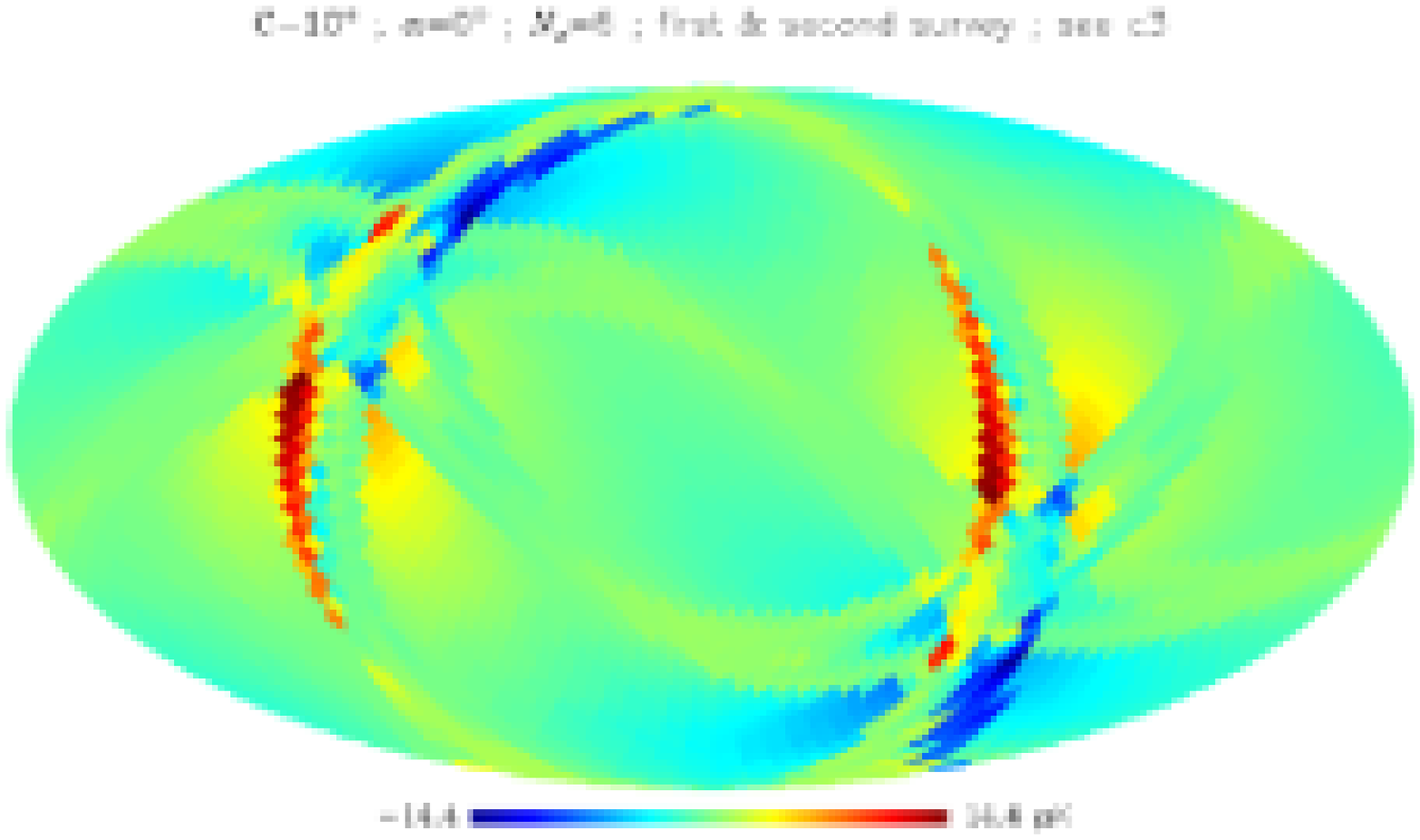}
   \end{tabular}
\caption{Maps of DSC
in the case of a cycloidal scanning strategy (C) 
with a semi-amplitude of $10^\circ$, $\alpha = 0^\circ$ or $45^\circ$ and for 
$N_c = 12$ or 6.
The parameters of each case are indicated 
above each map together with the reference to the panel in Figure~\ref{cl_allcases} where
the corresponding APS is displayed. 
Galactic coordinates and Mollweide projection are used.} 
\label{map_c}
\end{figure*}

\subsection{Numerical results}
\label{numericalresults}

We carried out the simulation described in the previous section
for various scanning strategies and configurations.

First, we considered the scanning strategy (the NSS) previously adopted in Section~\ref{model}
with the spin axis always in the antisolar direction 
for three choices of the angle $\alpha$: $0^\circ, 10^\circ$ and $45^\circ$.
The corresponding maps for some representative cases are shown in Figures~\ref{map_nss} and 
\ref{map_nss_c}. 
We considered the maps obtained coadding the TOD from each of the two surveys separately
and from the two surveys together. In the case with $\alpha = 0^\circ$ we considered also 
the maps obtained from one whole survey plus 
a second uncomplete (2/3) survey. 

We then considered a cycloidal scanning strategy (C) with periods of 
6 months (i.e. a number, $N_c$, of complete cycloids per 360 days equal to 2),
2 months (i.e. $N_c = 6$), and 1 month (i.e. $N_c = 12$). 
The corresponding maps for some representative cases are shown in Figures~\ref{map_nss_c} and 
\ref{map_c}. 
In the cases with $N_c = 2$, we considered $\alpha = 0^\circ$ and  $10^\circ$;
in the cases with $N_c = 12$, we considered $\alpha = 0^\circ$ and $45^\circ$;
in the cases with $N_c = 6$, we considered $\alpha = 0^\circ$. 
The semi-amplitude of the precession cone has been set to $10^\circ$.
For comparison, we considered also a semi-amplitude of $5^\circ$
in the case $\alpha = 0^\circ$ and $N_c = 12$.
Again, we analysed the cases of each of the two surveys 
separately and of the two surveys together.

We observe that it is preferable to first produce the map corresponding 
to each survey separately and then to construct the combined map from the two 
surveys as a simple average of the two separate maps. 
In fact, 
even a relatively small difference in the number of hits
per pixel in the two surveys
(generated by a
combination of pixel shape and effective observational strategy) 
can introduce a remarkable spurious unbalance in the average when it
is done by jointly treating the TOD from the two 
surveys~\footnote{Note also that considering a single survey,
the numerical error in the TOD quoted above is further dropped by
averaging. This is not strictly true by combining two surveys because
pixel shape and effective scanning strategy imply a not perfect geometrical 
symmetry between the TOD from the two surveys and the presence 
of (very small) residual deviations from the ideal case.}.

\begin{figure*}
   \begin{tabular}{cc}
\hskip -1.cm
\includegraphics[width=11.5cm]{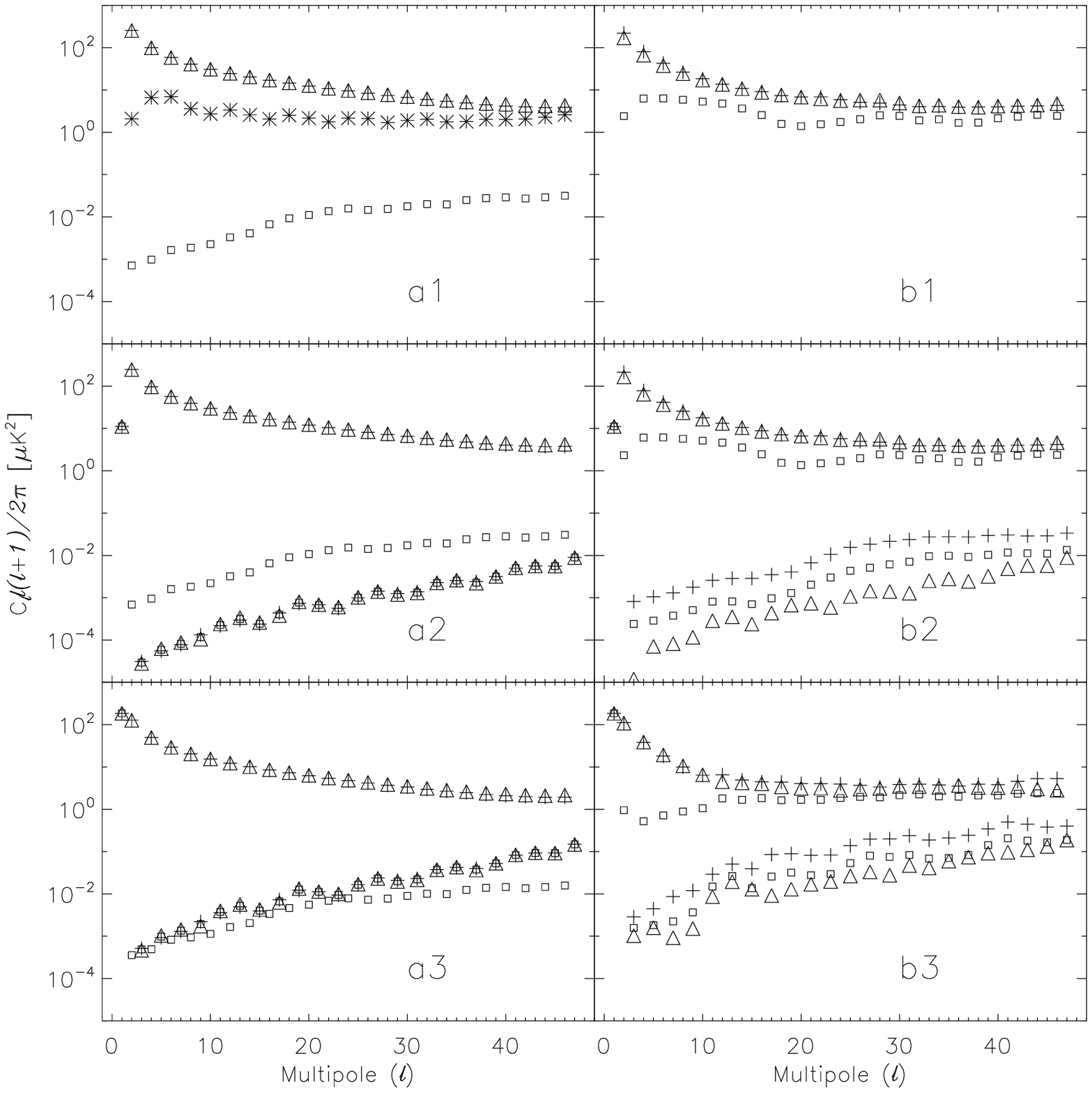}
\hskip -1.cm
\includegraphics[width=11.5cm]{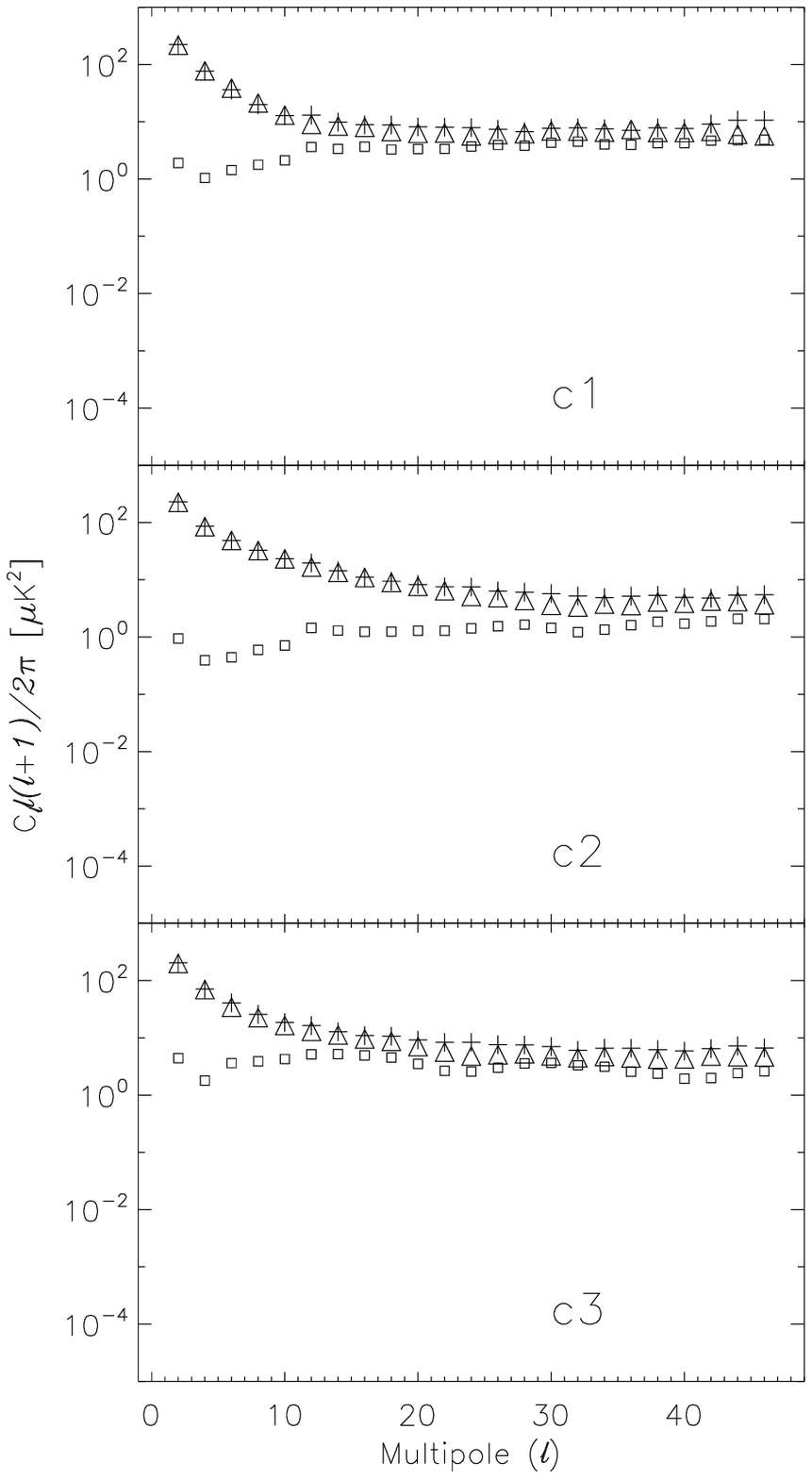}
   \end{tabular}
\vskip -1.5cm
\caption{APS of the DSC
for various configurations.
Panels a1, a2, a3 refer to the NSS, respectively for $\alpha = 0^\circ, 10^\circ$ and $45^\circ$. 
The others panels refer to cycloidal scanning strategies. 
Panels b1 and b2 refer to a 
semi-amplitude of $10^\circ$ and $N_c=2$, respectively for $\alpha = 0^\circ$ and $10^\circ$.
Panel b3 refers to a 
semi-amplitude of $10^\circ$ and $N_c=12$, and $\alpha = 45^\circ$.
Panels c1, c2, c3 refer to $\alpha = 0^\circ$ for a
semi-amplitude of $10^\circ$ and $N_c=12$ (panel c1), 
a semi-amplitude of $10^\circ$ and $N_c=6$ (panel c3),
a semi-amplitude of $5^\circ$ and $N_c=12$ (panel c2).
In all panels triangles and crosses refer to a single survey and 
squares to the combination of the two surveys. In panel a1, asterisks
refer to the case of one complete survey combined with an uncomplete
(2/3) survey. See also Figures~\ref{map_nss}, \ref{map_nss_c}, and \ref{map_c}. 
Note the scanning strategy dependence 
of suppression of the power at the even multipoles
in the case of the combination of the two surveys. 
Note that the power at odd multipoles
is different from zero only for $\alpha \ne 0$; this is of particular relevance 
in the case of $\ell =1$ (dipole), and may have implications also on
data calibration with the dipole and dipole modulation.
See also the text.} 
\label{cl_allcases}
\end{figure*}

The APS obtained analysing the various maps are 
reported in Figure~\ref{cl_allcases} for some representative cases.
In each of the maps in Figures~\ref{map_nss}, \ref{map_nss_c}, \ref{map_c} we report the 
corresponding panel in Figure~\ref{cl_allcases}.

\subsubsection{Simple scanning strategies}
\label{nss}

The simple view to the maps obtained in the 
case of the NSS confirms the results derived 
through the analytical approach.

Figure~\ref{map_nss} clearly shows the quadrupole produced by the dipole
straylight contamination when only a single survey is considered.
Note that the pattern is symmetrical in the two surveys,
as expected.

Considering the average of two surveys, the resulting map
is essentially zero, except for a small ring, corresponding to the 
edges of the surveys, where it is in practice
impossible to exactly balance the signal from the two surveys
as a result of the combination of pixel shape and effective 
observational strategy. 

If one of the two surveys is not complete (see Figure~\ref{map_nss}),
a remarkable quadrupole again appears. 
Figure~\ref{cl_allcases} clearly shows this effect in terms of APS.

Note that, as expected from the analytical treatment
for $\alpha = 0$,
our computation gives an APS different from zero only 
for the even multipoles.

Note also the power decreasing for increasing multipoles.
At low multipoles, we have checked that the APS found numerically 
agrees with the analytical prescription. 

Differently from the ideal analytical case,
the combination of two complete surveys implies 
a non-vanishing APS because of small deviations of the exact 
balance found analytically. In any case, these effect
(that is found to decrease by increasing the resolution of 
the simulation) is absolutely negligible in practice 
in this context.

Much more remarkable is the APS found when one of the 
two surveys is not complete. 
In practice, this may be the case in realistic 
experiments (even for the most symmetrical NSS) because of
the different location of the various receivers 
on the focal plane.
A particular care to this aspect should then be taken 
in the analysis of low multipoles.

As analytically expected, for $\alpha \ne 0$
a non-negligible dipole contribution also appears, as evident in the map
in Figure~\ref{map_nss_c}.
It has the same amplitude in the two surveys, as predicted.
Note that in this case the APS is different from zero
also for the odd multipoles, in agreement with 
the analytical results (see Figure~\ref{cl_allcases}). Clearly, the effect is in any case
very low, appearing only to higher order in $\alpha$.

For relevant values of $\alpha$, a case that cannot be exhaustively treated
through the analytical approach,
the power of the dipole
and of the other odd multipoles significantly increases
(see Figures~\ref{map_nss_c} and \ref{cl_allcases}). Note also in Figure~\ref{map_nss_c} the presence 
of remarkable 
features introduced by the significant angle between
the directions of the main spillover and of the spacecraft spin axis.

We note that an approximation for the scaling with $\alpha$ 
of the APS with $\ell =1$ can be
derived by the analytical expressions in equations (\ref{T10}) and (\ref{T1pm1}),
holding in the limit of small values $\alpha$, replacing $\alpha$ 
with $\sin \alpha$ which implies the $\sin^2 \alpha$ dependence
evident in panels a2 and a3 of Figure~\ref{cl_allcases}.
Note also that this dipole scaling holds almost 
independently of the details of the scanning strategy as clear
from panels b2 and b3 of Figure~\ref{cl_allcases} (see also the next subsection), 
because of its large angular scale origin.

\subsubsection{Cycloidal scanning strategies}
\label{css}

The complication of the observational strategy implied by a cycloidal
option has a clear impact both 
on the final maps and on the APS.

We first consider slow precessions (6 month period, $N_c =2$).
In the case $\alpha=0^\circ$, 
the most relevant difference with respect to the case of the NSS appears in the 
combination of the two surveys. As evident in Figure~\ref{map_nss_c}, structures appear
on various angular scales because of the violation of the balance
between the TOD of the two surveys implied by this observational strategy.
The APS found in this case (see Figure~\ref{cl_allcases}) is similar to
that found in the case of the NSS but in the absence of 
completeness of one of the two surveys. 

In the case $\alpha=10^\circ$ we find very similar results, except
for the expected presence of the dipole and of the other odd multipole terms.

Faster precessions (1 month period) with a significant value of $\alpha$ ($45^\circ$,
see Figures~\ref{map_c} and \ref{cl_allcases}), imply a remarkable dipole signature, analytically 
expected,
a relatively more efficient smoothing of some large scale features 
in the combination of the two surveys, 
and then a relative decreasing of the power at multipoles less than $\sim 10$
with respect to the previous case. 

Keeping $\alpha = 0^\circ$, we exploit the impact of 
different precession periods (see Figures~\ref{map_c} and \ref{cl_allcases}).
Note in Figure~\ref{map_c} how the precession period appears in the maps as a ``finger print'';
this effect is evident in general for $N_c > 2$.
This results is analogous to that found for the number of hits 
in scanning strategy analyses \citep{DupacTauber05},
because different precession periods imply different periodic modulations
in the dipole amplitude as observed by the main spillover.
The comparison between panels b1, c1, c3 in Figure~\ref{cl_allcases} shows that,
except for the quadrupole, the first weak and broad low $\ell$ bump in the APS
(properly, in terms of $C_\ell \ell (\ell+1) / 2\pi$) 
migrates towards higher multipoles for increasing $N_c$. This is 
a direct effect of the decreasing of the angular scale of the ``finger print''
structures for decreasing precession periods.

Finally, we have investigated on the impact of the precession angle amplitude
by considering also an angle of $5^\circ$ (panel c2 in Figure~\ref{cl_allcases}), 
instead $10^\circ$.
We find a certain decreasing of the power at multipoles less than 20--30 with respect 
to the previous cases, in particular jointly considering the 
two surveys. This can be understood on the basis of a simple continuity argument:
decreasing the amplitude, the cycloidal scanning strategy tends to the NSS.

\subsubsection{Symmetrical main spillover}
\label{sms}

For sake of completeness, we have carried out also some simulations
assuming a far sidelobe response azimuthally symmetrical with respect to the direction
of the main beam~\footnote{This scheme is inspired by the case of
observations taken with an antenna pattern similar to that of 
a feedhorn non-coupled to a telescope.}. 
We assume again a far sidelobe with a Gaussian 
profile, but in this case this refers only to the dependence on the colatitude 
from the beam centre axis.
We assume again FWHM$_{ms} = 20^\circ$, but it refers here only to azimuthal cuts.
We renormalize the maximum sidelobe response to have again a given ratio $p$ (=~0.01
in the current computations)
between the sidelobe response integrated over the whole solid angle and
the dominant main beam integrated response. In the simulations, we consider 
two cases, with the maximum sidelobe response
located at an angle of $90^\circ$ from the main beam 
(i.e. $\alpha = 0^\circ$) or at an angle of $45^\circ$ from the main beam
(i.e. $\alpha = 45^\circ$).

We considered again the NSS and cycloidal modulations of the spin axis, in this case
with $N_c=12$ and a semi-amplitude of $10^\circ$.

As expected on the basis of an obvious symmetry argument, 
in the case $\alpha = 0^\circ$ we find no effect independently of the
adopted scanning strategy.

For increasing $\alpha$, the above full symmetry is broken and in fact 
in the case $\alpha = 45^\circ$ we find a non-vanishing effect. On the other hand,
the azimuthal symmetry adopted for the far sidelobes implies a strong suppression
of the power at multipoles $\ell > 1$ and the only surviving relevant term 
is the dipole~\footnote{In this case, the far sidelobe azimuthal symmetry, significantly different 
from that considered in the previous cases, reflects into a suppression of the power at 
even multipoles and into a presence of a very weak power at odd multipoles, 
in  any case smaller than 
$\sim 0.3 \mu$K$^2$ (and of $\sim 0.01-0.1 \mu$K$^2$ for $\ell \lsim 9$) 
in terms of $C_\ell \ell (\ell+1)/ 2\pi$, 
because of high order terms in $\alpha$, analogously 
to the previous case (again the numerical estimates refer to $p=0.01$).}, as 
can easily understood by considering that for the case
$\alpha \rightarrow 90^\circ$ we recover the case of the original dipole 
pattern, smoothed and decreased of a factor $p$.
The simulations carried out in these cases for $\alpha = 45^\circ$ show a 
power at $\ell = 1$ very close to that found in the previous subsections
for $\alpha = 45^\circ$ for the same value of $p$.

\begin{figure*}
   \begin{tabular}{ccc}
\includegraphics[width=3.2cm,angle=90]{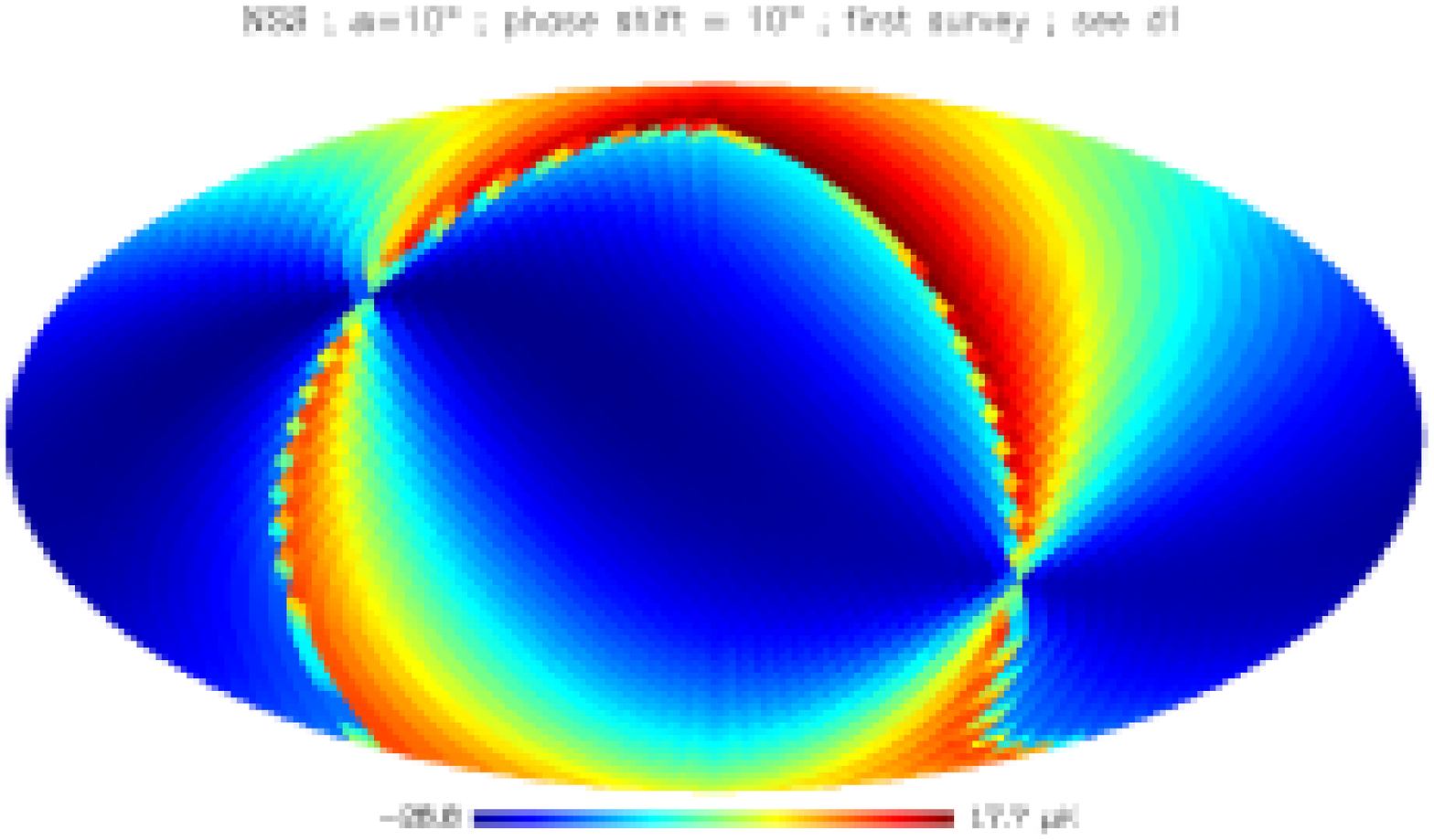}&
\includegraphics[width=3.2cm,angle=90]{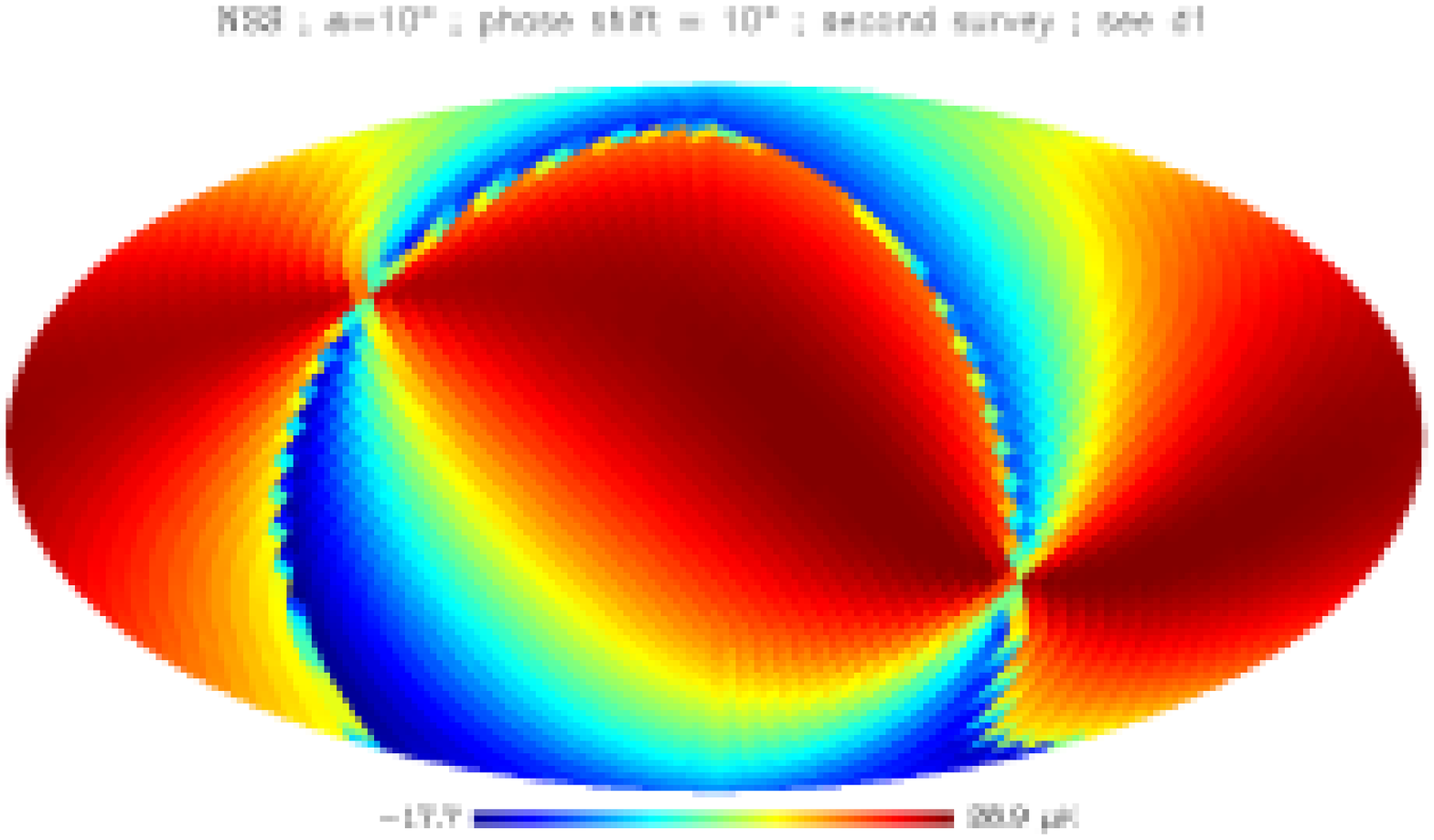}&
\includegraphics[width=3.2cm,angle=90]{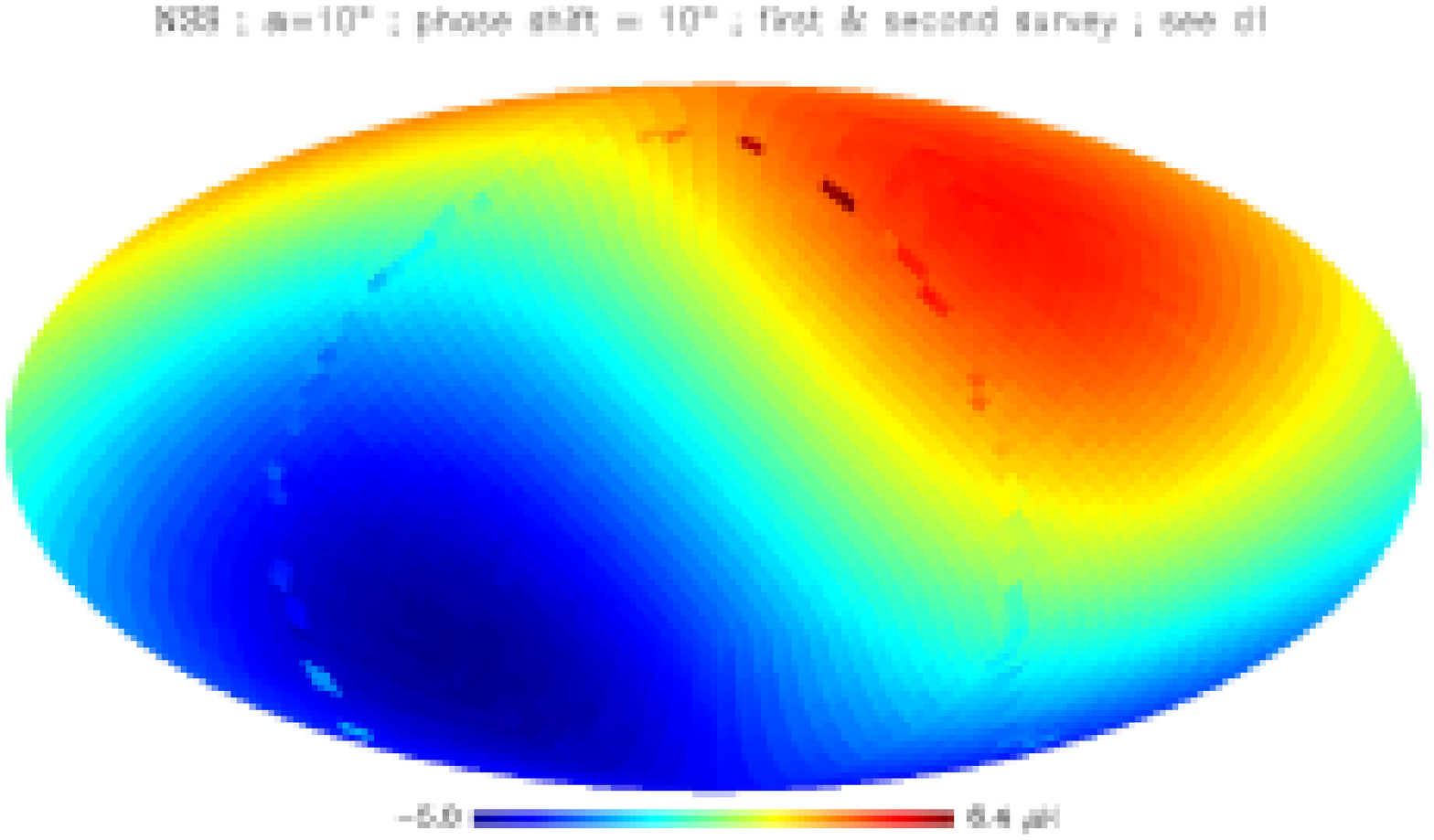}\\
\includegraphics[width=3.2cm,angle=90]{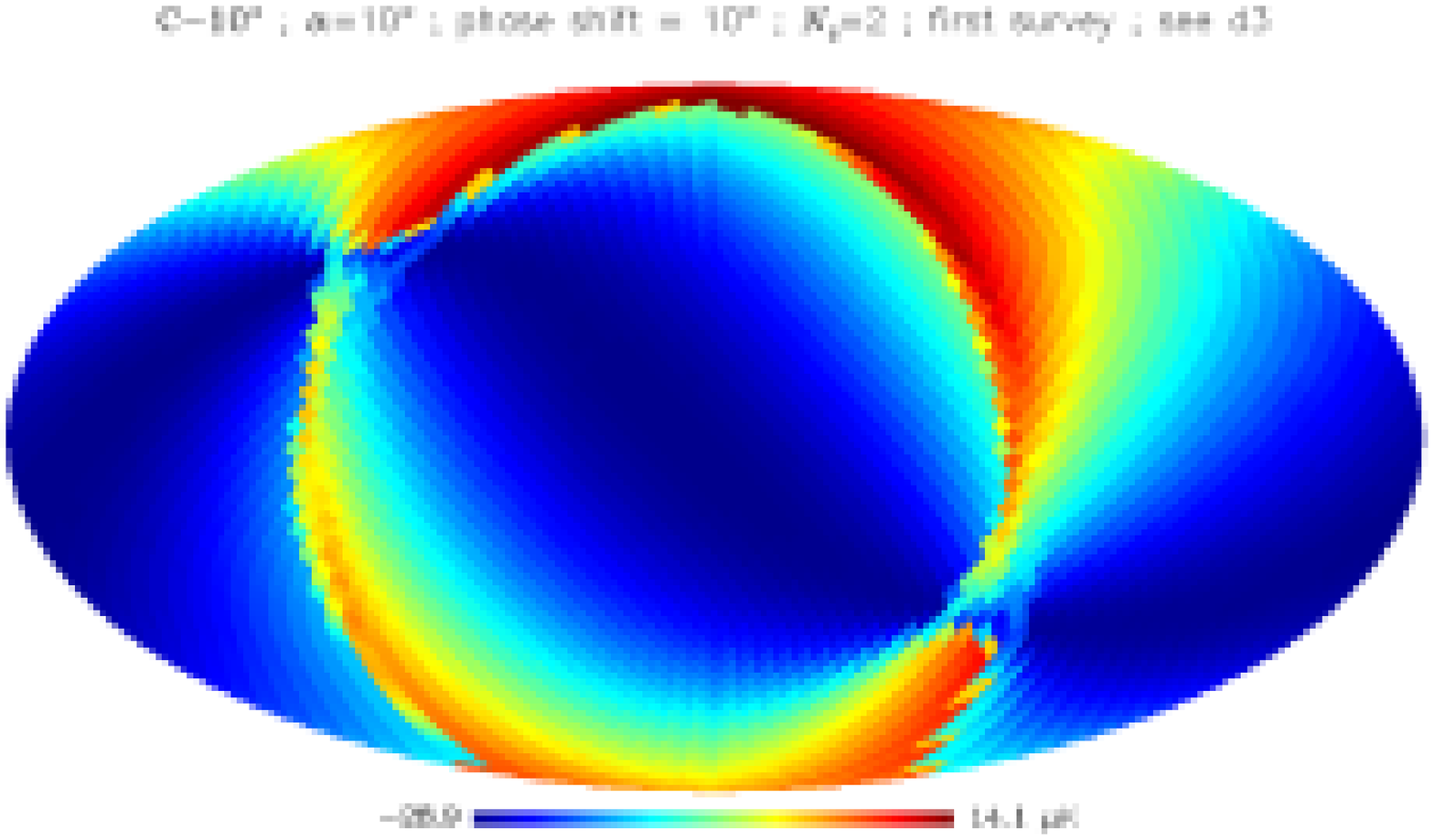}&
\includegraphics[width=3.2cm,angle=90]{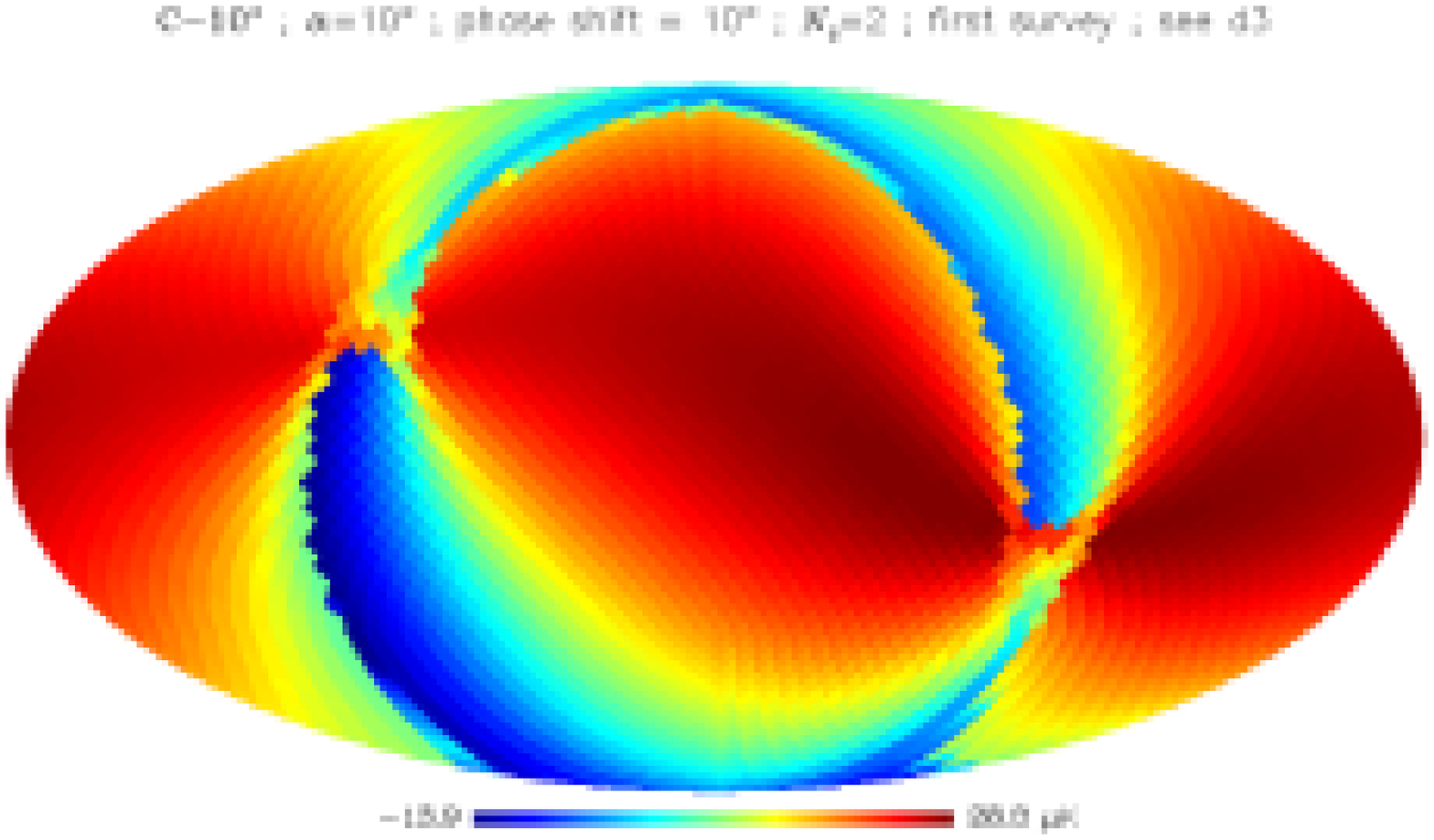}&
\includegraphics[width=3.2cm,angle=90]{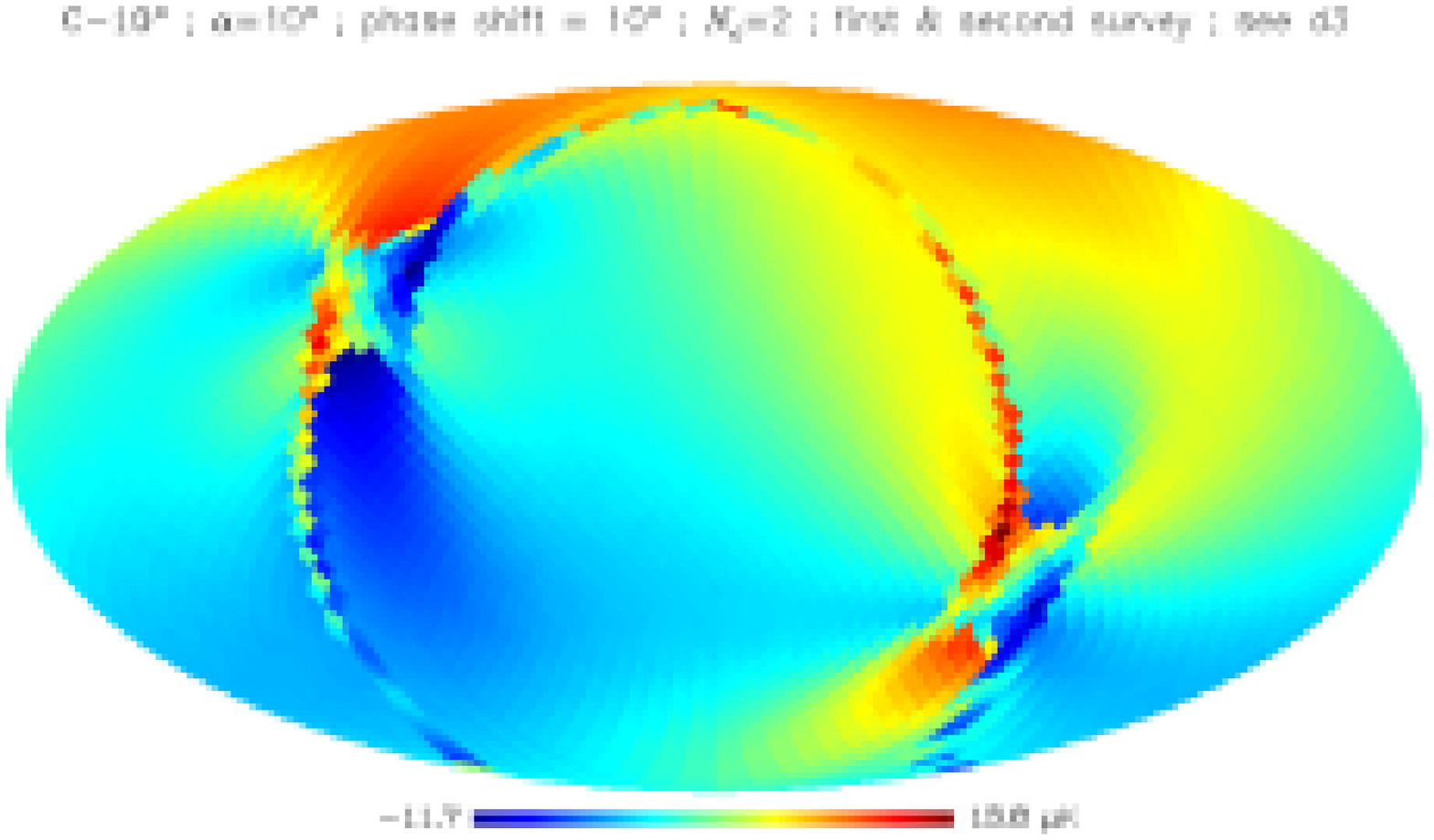}\\
\includegraphics[width=3.2cm,angle=90]{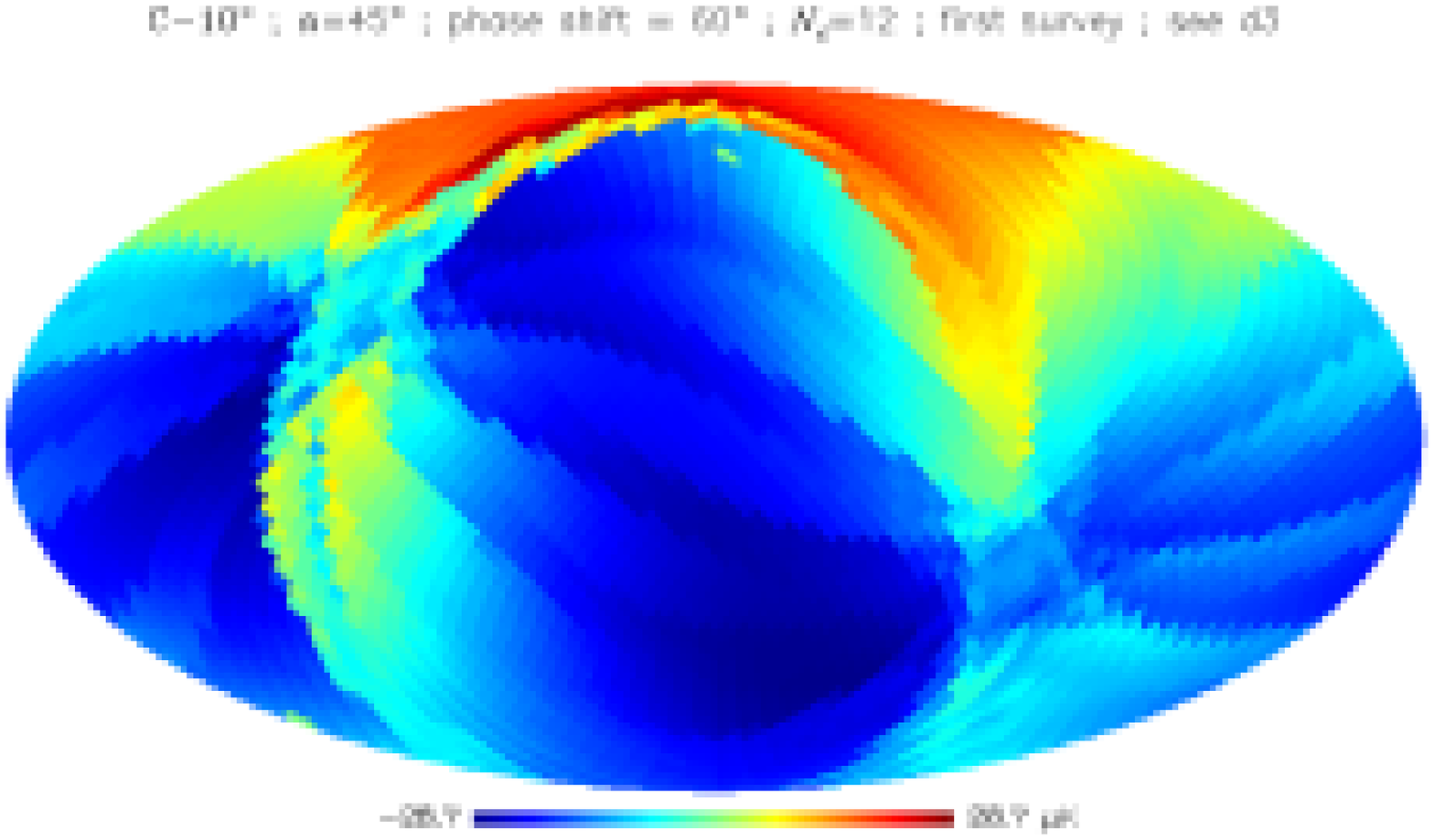}&
\includegraphics[width=3.2cm,angle=90]{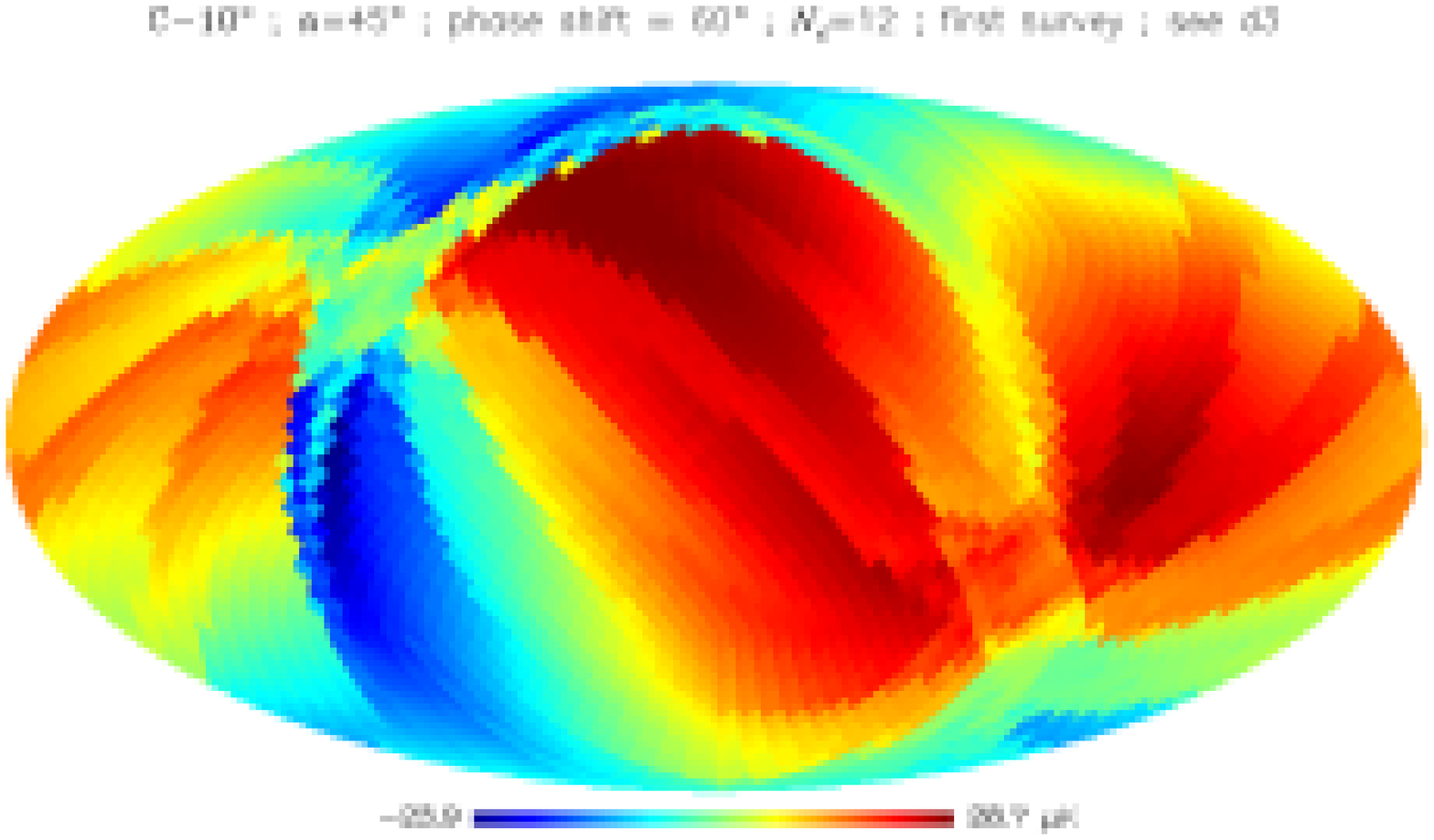}&
\includegraphics[width=3.2cm,angle=90]{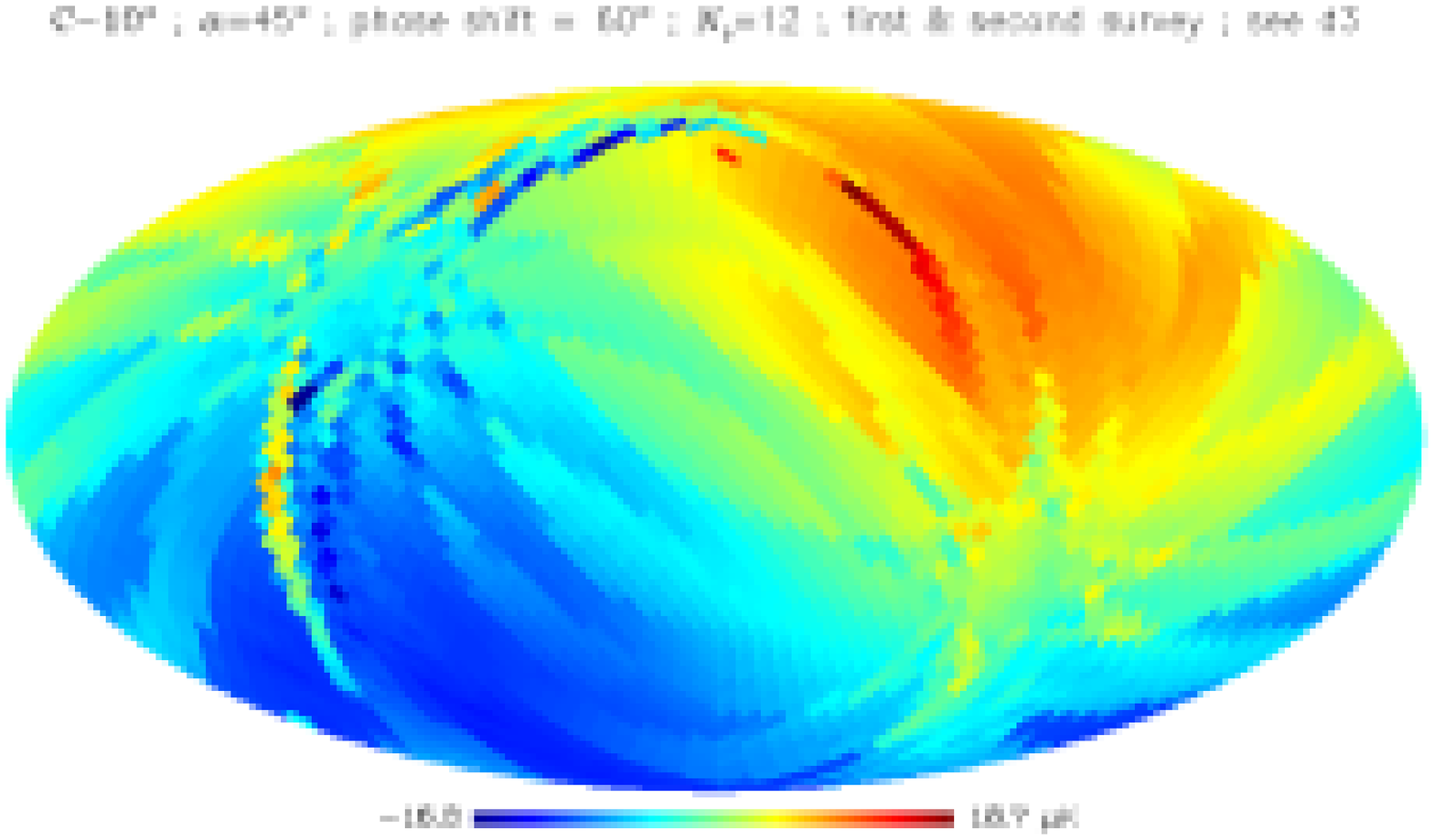}
   \end{tabular}
\caption{Maps of DSC in the case of the NSS for small values of $\alpha$ and of the phase shift
(top panels)
and in the case
of a cycloidal scanning strategy with small (middle panels) and
large (bottom panels) values of $\alpha$ and of the phase shift.
The parameters of each case are indicated
above each map together with the reference to the panel in
Figure~\ref{cl_allcases_shift}
where the corresponding APS is displayed.
Galactic coordinates and Mollweide projection are used.} 
\label{map_shift}
\end{figure*}

\section{Displacement of the main spillover}
\label{displacement}

The relaxation of the assumption of main spillover centre location on the
plane defined by the spin axis and the telescope line of sight
can be easily treated through numerical simulations. Because of the spacecraft
rotation about the spin axis,
any displacement of the main spillover centre from that plane
is geometrically equivalent to a proper phase shift
(constant for all scan circles) between
the TOD containing the main beam centre pointing directions and the
TOD containing the main spillover straylight signals.
The TODs from the simulations described in the
previous section can be then easily reordered to describe this case.

\begin{figure}
\includegraphics[width=11.5cm]{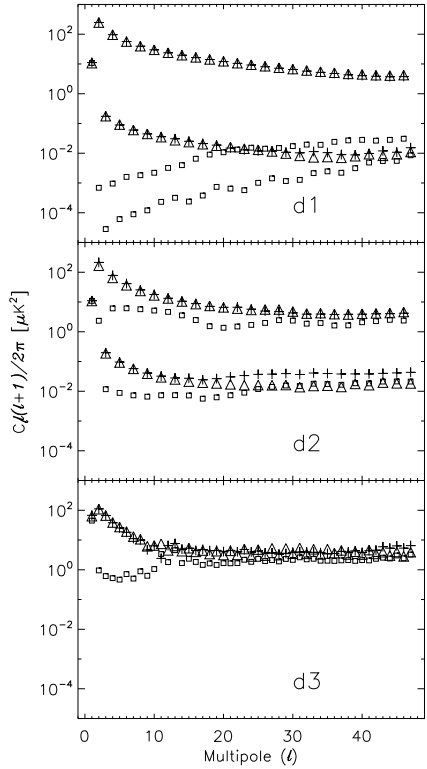}
\vskip -1.5cm
\caption{APS of the DSC
for various configurations.
Panel d1 refers to the NSS for $\alpha = 10^\circ$ and a phase shift of $10^\circ$.
The others panels refer to cycloidal scanning strategies with a
semi-amplitude of $10^\circ$.
Panel d2 refers to $N_c=2$, $\alpha = 10^\circ$ and a phase shift of $10^\circ$.
Panel d3 refers to $N_c=12$, $\alpha = 45^\circ$ and a phase shift of $60^\circ$.
In all panels triangles and crosses refer to a single survey and
squares to the combination of the two surveys.
Compare panel d1, d2, and d3 respectively with panel a2, b2, and b3 in
Figure~\ref{cl_allcases}.
See also Figure~\ref{map_shift}.
Note the significant power at odd multipoles in the panel d3 due to the relevant
values of $\alpha$ and of the phase shift.
See also the text.}
\label{cl_allcases_shift}
\end{figure}

Figure~\ref{map_shift} shows the maps obtained (obviously assuming $\alpha \ne 0$)
for some representative choices of the relevant parameters
while Figure~\ref{cl_allcases_shift} displays the corresponding APS
(for instance, the phase shift refers here a ``delay'' 
of the main spillover straylight signals).

Note that, as expected, in the case of small values of $\alpha$ and of the
phase shift the results do not significantly change
(see panels d1 and d2 in Figure~\ref{cl_allcases_shift})
with respect to the case
in which the main spillover centre is located on the
plane defined by the spin axis and the telescope line of sight
because the leading terms are still those at the lowest order in $\alpha$
and in the phase shift, as discussed above.
On the contrary, for significant values of $\alpha$ and of the phase shift
the odd multipole power (not only at $\ell = 1$)
is similar to that found for the contiguous even multipoles
(see panel d3 in Figure~\ref{cl_allcases_shift})
for both a single survey and the average of the 
two surveys~\footnote{In the case of small $\alpha$ and significant
phase shift 
we numerically find again that the power at a given even multipole
is larger
than that at the contiguous odd multipoles
(again except at $\ell = 1$), although with a
mitigation of its dominance, depending on the specific
set of considered parameters.}.

Analytically this phase shift can be parametrized generalizing equations (\ref{thetamsexp}) and
(\ref{varphimsexp}) as follows
\begin{eqnarray} 
& & \theta_{ms} = {\pi \over 2} - \cos (\theta _{mb} \mp \beta) \, \alpha + {\cal O} \left( \alpha^3 \right)
\, , \label{thetamsexpwithbeta}\\
& & \varphi_{ms} = \varphi_s \pm \sin (\theta _{mb} \mp \beta) \, \alpha + {\cal O} \left( \alpha^3 \right)
\, \label{varphimsexpwithbeta},
\end{eqnarray}
with $\beta$ defining the displacement from $\theta_{mb}$ and where
upper (lower) signs refer to North (South) towards South (North) motion of the main beam.
Of course, when $\beta$ is vanishing, equations (\ref{thetamsexp}) and (\ref{varphimsexp}) are recovered. 

Replacing (\ref{thetamsexpwithbeta}) and (\ref{varphimsexpwithbeta}) in the definition
of $I_{SL}$ [see equation (\ref{ISL1survey}) or (\ref{ISL2survey})] it is possible to compute the map
\be
T_{\ell m}^{SL} = T_{\ell m}^{0} + \alpha T_{\ell m}^{1}(\beta) \, .
\ee 
Here $T_{\ell m}^{0}$ is given by equation (\ref{intonesurvey}) [i.e. 
the introduction of $\beta$ leaves uneffected the zeroth order, see equations (\ref{thetamsexpwithbeta}) and (\ref{varphimsexpwithbeta})] while the first 
order is changed as
\be
T_{\ell m}^{1}(\beta) = \cos \beta \, T_{\ell m}^{\parallel} + \sin \beta \, T_{\ell m}^{\perp}
\, ,
\ee
where label $^{\parallel}$ and $^{\perp}$ refer to the plane defined by the directions of
the spin axis and main beam centre. 
The coefficient $T_{\ell m}^{\parallel}$ is given by equation (\ref{Tlm1storder})
and $T_{\ell m}^{\perp}$ is 
\be
T_{\ell m}^{\perp} = 2 c_1 \left[ -1 + (-1)^m \right] I_1 
- c_{23} \left[ 1 + (-1)^m \right] I_2
\, ,
\label{perp}
\ee
where $c_{23}=c_2 + c_3$, and
\be
I_1 = \int_{-1}^{1} \!\!\!\! dx \int_0^{\pi} \!\!\!\! 
d \varphi \sqrt{1-x^2} \, Y_{\ell m}^{\star}(\theta(x), \varphi)
\, ,
\label{integr_i1}
\ee
and
\be
I_2 = \int_{-1}^{1} \!\!\!\!dx \int_0^{\pi} \!\!\!\! d \varphi 
(d_1 \cos \varphi - d_2 \sin \varphi) x Y_{\ell m}^{\star}(\theta(x), \varphi) 
\, ,
\label{integr_i2}
\ee
with $\theta(x) = \arccos x$.
For symmetry reasons in equation (\ref{perp}) only odd $\ell$ contributions turn on.
When $m$ is odd then $I_2 =0$ (because of the integration over $x$) and $I_1$ can be 
non-vanishing 
(for $\ell > 1$, $m$ has to be different from $\pm 1$);
when $m$ is even then $I_1 =0$ (because of the integration over $x$) and $I_2$ 
can be non-vanishing (for $\ell > 1$, $m$ has to be different from $0$).
For $\ell =1$, we compute
\be
T_{10}^{\perp} = {8 \over 3} \sqrt{{3\over{4 \pi}}} c_{23} d_2
\, ,
\ee
and
\be
T_{1\pm 1}^{\perp} = {16 \over 3} \sqrt{{3\over{2 \pi}}} c_1 i
\, .
\ee
We conclude that to first order in $\alpha$ only odd $\ell $ are present 
and the shift $\beta$ modifies the dipole term and switch on $\ell > 1$ terms.
When $\beta \sim \alpha$ than the shift is a second order term and to linear order
does not appear. 
This is in agreement with the numerical result found
in the linear regime in $\alpha$, 
as evident from the
comparison of panel a2 
of Figure~\ref{cl_allcases} with panel
d1 of Figure~\ref{cl_allcases_shift} (see also footnote 15).

Finally, we give the average $\hat T_{\ell m}$ over two or an even
number of surveys in case of a shift $\beta$ different from zero:
\be
\hat T_{\ell m} = \alpha \, \cos \beta \, T_{\ell m}^{\parallel}
\, .
\ee
Since, as already mentioned, $T_{\ell m}^{\parallel} \neq 0$ only for $\ell = 1$
(see equation (\ref{Tlm1storder}) and Subsection \ref{survivingsystematic}; see also
Subsection \ref{resultsoddnumbersurvey}),
only the dipole term survives after two (or an even number of) surveys,
also for $\beta \neq 0$. 
This result agrees with panel d1 of Figure~\ref{cl_allcases_shift} where $\beta \sim \alpha$.

\section{Implications for spinning space missions}
\label{numremarks}

Numerical simulations confirm the analytical result that this effect 
is particularly remarkable in the case of an odd number of surveys,
while a proper average of an even number of surveys greatly reduces its
amplitude.

On the other hand, for complex scanning strategies and or in the presence
of uncompleteness of one of the considered surveys the suppression through
averaging of this effect is significantly reduced.
Clearly, the choice of a given scanning strategy is driven by many other
aspects (mission constraints, sky coverage, redundancy, overall reduction
of systematic effects, and so on; see e.g. \citet{bernard}, \citet{DupacTauber05}, 
\citet{maris}).
Our analysis shows that, in particular at low multipoles, a special care
should be taken in combining data from multiple surveys.
Since, typically, at low $\ell$ the sensitivity is not a problem 
in comparison to systematic effects and foreground removal,
it could be preferable to avoid to include in the low $\ell$ analysis
data exceeding an even number of complete surveys to reduce a priori
the amplitude of this effect, or, at least, to compare the results found 
in this way with those derived by using the whole set of data.

We have considered here the 
DSC in the case of 
a single receiver.
This is the case of {\sc Planck}-like optical configurations in which 
the sky signal is compared with a reference signal.
In other optical configurations, for example in the case of WMAP, 
CMB anisotropies are mapped by comparing the sky 
signal in two different sky directions. Clearly, in this case 
the effective DSC
comes from
the difference between the dipole signal as seen by the far sidelobes 
corresponding to the considered pair of receivers.
Therefore, the final effect will mainly depend on the different orientation 
and response of their main spillovers
\footnote{For example, if they simultaneously point to
the same sky region with similar responses the final effect
will be significantly attenuated.
On the contrary, if they simultaneously point to sky regions
with opposite dipole signs, the effect will be amplified.}.

\subsection{Calibration}
\label{calib}

We note that for $\alpha \ne 0$,
and according to the value of $p$, 
a non-negligible power at $\ell = 1$ may appear.
It increases with $\alpha$ and can produce on the map
peak values of $\sim \pm 10 \mu$K (for $\alpha$ of few tens
of degrees) for $p=0.01$. 
Clearly, the value is small compared to the kinematic dipole 
amplitude of $\sim 3$~mK, but, depending on $p$, it could be non-fully negligible 
when compared with the calibration accuracy of CMB space mission with both dipole 
and its $\sim 0.3$~mK
modulation during the year induced by the Earth 
motion around the Sun \citep{bersanelli97,piat02,cappellini03,jarosik06}.

\subsection{Removal during data analysis}
\label{removalduringdataanalysis}

Clearly, provided that the far pattern is well known, 
it is possible to accurately simulate this effect for the mission effective
scanning strategy
or to apply deconvolution map making schemes working on the full 
beam pattern \citep{harrison06}
in order to subtract the DSC from data, maps, and APS.
The first strategy has been in fact adopted by the WMAP team
to remove the overall 
(i.e. from the global signal sum of the various components)
straylight contamination 
for all the frequency channels during the data analysis of the three
year data \citep{jarosik06}.

Clearly, a precise subtraction of the DSC
relies on a very accurate knowledge of the far beam.

For example, an error produced by an overall underestimation (or overestimation) 
of the main spillover response
of 3~dB (or 2~dB, or 1~dB) is equivalent in this simple scheme
to a $\simeq 100$\% (resp., 60\% or 26\%) 
error on the value of $p$ (or, equivalently, of 
$f_{SL}$, $F_{SL}$, see equation (\ref{FSL})). The residual contamination in the map
and in $\sqrt{C_\ell}$ can be obviously obtained by the results presented
above rescaled by the same factor. Analogously, 
Figures~\ref{fig3} and \ref{fig4} can be used
to understand the error on the quadrupole recovery.
While for $B < 0$, even a poor knowledge of $F_{SL}$ (i.e. of $p$)
implies in any case at least a partial removal of this systematic 
effect  (with respect to the case in which this subtraction is not applied to the data
-- obvioulsy, this subtraction has a meaning only for $\Delta p /p < 100$\%),
the treatment of the case $B > 0$ is more difficult because the final effect of 
this removal will depend on the true value of $F_{SL}$ (or $p$), 
on the error in its knowledge, and on the intrinsic quadrupole $C_2^{SKY}$, given the
parabolic behaviour of $y(F_{SL})$. This calls for a particular care
in the knowledge of the main spillover, at a level better than 
$\simeq 1$~dB.

\subsection{Comparison with Galactic straylight contamination}
\label{comparison}

\citet{burigana2} presented an analysis of the Galactic
straylight contamination at 100~GHz and 30~GHz taking into account
the optical simulation by \citet{sandri}, in the context
of {\sc Planck} LFI optimization work. 

For values of $p \approx 0.001$ (see the quantity $f\%$ in Table~3 in \citet{sandri}),
at 100~GHz \citet{burigana2} (see \S~5) found 
$C_\ell \ell (\ell +1) /2\pi \lsim 10^{-2} \mu$K$^2$.
Rescaling the values reported in Figure~\ref{cl_allcases} for $p = 0.01$ to the case 
of $p \simeq 0.001$ we find 
$C_\ell \ell (\ell +1) /2\pi$ in the range $\simeq [3 - 0.03] \mu$K$^2$
(or in the range $\simeq [0.1 - 0.01] \mu$K$^2$)
from lower to higher multipoles
for the case of a single survey (or of the combination of two surveys).

We then conclude that, at least at frequencies close to the minimum of Galactic
foregrounds (70~GHz), the DSC is 
more relevant at low multipoles than the Galactic straylight contamination.

Clearly, the different frequency scaling of CMB and Galactic foregrounds
implies an opposite conclusion at significantly lower and higher 
frequencies \citep{dezotti99,bouchetgispert99}.

Finally, we observe that the APS of Galactic straylight contamination
in the case of {\sc Planck} LFI
also shows a power typically larger for the even multipoles than for the odd ones
- see Figure~9 of \cite{burigana1}-,
analogously to what expected from our analysis in the presence 
of a large dipole term in the considered diffuse component 
for a far beam dominated by the main spillover
feature not far from the spin axis.

\subsection{Polarization measurements}
\label{implicationforpolirization}

The CMB polarization anisotropy is typically observed by 
combining the signals from a set of polarization receivers.
For example, for differential radiometers as in {\sc Planck} LFI
or in WMAP the minimum set of receivers for polarization 
observations consists of four radiometers coupled to two feeds.
Although the CMB dipole is not polarized, the 
DSC may affect polarization measurements because of differences
in the intensity in the various receivers and,
in the simple case of combination of data from four radiometers coupled 
to two feeds from the differences in the dipole straylight signals in each pair
of radiometers associated to the same feed.
Therefore, the method described in \S~4 of 
\citet{burigana2}
can be applied here to provide simple numerical estimates.
Clearly, also in this case the effective 
DSC will mainly depend on the different orientation 
and response of the main spillovers corresponding to each pair
of radiometers associated to the same feed.
Assuming a difference of a factor of $\simeq 2$ 
between the corresponding signals in these two radiometers, we find
that the contamination in the $Q$ and $U$ Stokes parameters
is similar to that present in the intensity.

This underlines the relevance of a very stringent control
of this systematic effect for future accurate polarization measurements,
being the CMB polarization $E$ mode at low multipoles less
than few~$\times 10^{-1} \mu$K (in terms of $(C_\ell \ell (\ell +1) /2\pi)^{1/2}$)
and being the expected $B$ mode smaller at least of a factor of two or three,
according to the current WMAP upper limits on the tensor to scalar ratio
\citep{spergel06}.

\section{Statistical analysis of the quadrupole}
\label{statistic}
The low multipoles of the CMB anisotropy
pattern probes the largest scales of our universe, far beyond the present
Hubble radius. The two experiments so far capable to measure such low 
multipoles - COBE/DMR and WMAP - have detected an amplitude for the 
quadrupole surprisingly low compared to theoretical expectations. It is 
not clear if the origin of this anomaly is statistically significant \citep{efstathiou03}, 
or related to foregrounds \citep{tegmark03,abramo,copi06} or 
possible systematic effects like the one discussed here. 
In the following we shall discuss the impact of DSC on this issue.

As discussed in Section~\ref{numremarks},
our modelling should be modified in 
order to investigate the DSC in the context of instruments 
based on differential measurements for CMB anisotropies, 
such as COBE/DMR and WMAP, which is beyond the scope of this work. 
However, we can make some considerations 
on the possible maximum 
\footnote{To this aim, we consider the case of an odd number of surveys.}
amplitude of this effect at low multipoles {\em as if} measured by an instrument 
like the one we have modelled.

First, we note that the value of the octupole has changed significantly 
from WMAP 1-yr release to the 3-yr one, reducing the problem of 
unexpected low amplitudes mainly to the quadrupole. Therefore, a 
systematic as DSC which leaves unalterated the octupole is in principle in better 
shape after the WMAP 3-yr release.

Second, as it can be seen from Section 4, the amplitude and correlation of 
the DSC contamination can in principle reduce {\em for the instrumental 
values 
chosen (which are realistic for {\sc Planck})} a theoretical expectation of the 
quadrupole ($C_2 \sim 10^3 \mu {\rm K}^2$) to the one observed. 
%
Therefore, it is important
to address the probability of how DSC affects the 
observed quadrupole, since such a systematic can increase or 
decrease $C_2^{\rm obs}$
depending on the sky realization (i.e. $T_{2 m}^{SKY}$). 

\begin{figure}
\centering
\includegraphics[width=6.7cm]{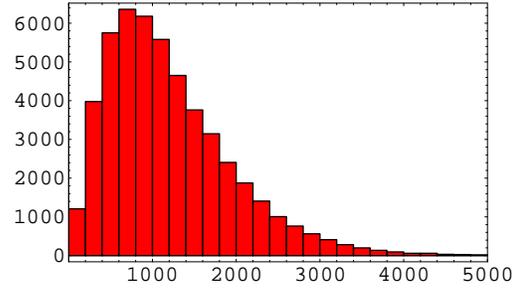}
\caption{Distribution (in terms of counts; $y$-axis) of the observed quadrupole $C_2$ 
($x$-axis).}
\label{fig9}
\end{figure}

\begin{figure}
\centering
\includegraphics[width=6.7cm]{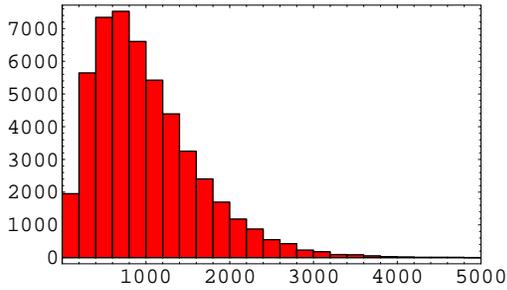}
\caption{Distribution (in terms of counts; $y$-axis)
of the intrinsic quadrupole $C_2^{SKY}$ ($x$-axis).}
\label{fig10}
\end{figure}

\begin{figure}
\centering
\includegraphics[width=6.7cm]{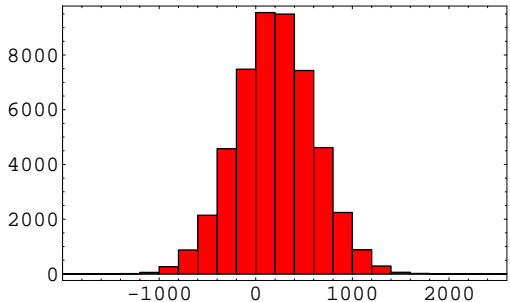}
\caption{Distribution (in terms of counts; $y$-axis) of the pure DSC 
$C_2-C_2^{SKY}$ ($x$-axis).}\label{fig11}
\end{figure}

We perform here a statistical analysis aimed at the computation of the probability 
of increase or decrease of the quadrupole (see \citet{ischia2006} for a preliminary analysis).
We have made $50000$ Gaussian distributed estractions of 
$T_{2 m}^{SKY}$ such that on average
$C_{2}^{SKY} \sim \, 10^3 \, \mu {\rm K}^2$ and we have computed the distribution for the observed 
quadrupole $C_2$ (see Figure~\ref{fig9}).  
In Figure~\ref{fig10} we show the distribution for the intrinsic quadrupole 
(i.e. $C_{2}^{SKY}$) and in Figure~\ref{fig11} we plot the distribution 
for the pure DSC (i.e. the difference $C_2-C_{2}^{SKY}$).
In Figure~\ref{fig11} we see that the distribution of DSC has a Gaussian profile 
with a mean different from zero, as can be understood from  
equations (\ref{c2sl_stat}) and (\ref{C2SKYSL}).
We check that the found mean ($202.3 ~\mu {\rm K}^2$) and
standard deviation ($404.3 ~\mu {\rm K}^2$)
of this distribution are
in agreement with the values expected respectively 
from equation (\ref{c2sl_stat}) and equation (\ref{C2SKYSL}),
for the chosen value of $F_{SL}$ ($=4.59\times 10^{-3}$).  
The probability of quadrupole amplitude increasing (decreasing)
is $69 \%$ ($31 \%$).

\section{Conclusion}
\label{conclusion}

We have developed an analytical model and numerical analyses to evaluate the
DSC in spinning CMB anisotropy missions.
Although our study is mainly devoted to the {\sc Planck} project,
the formalism and method are relatively general and allow to focus on
the most relevant DSC implications. 
We quantify this systematic effect
as a function of few parameters: the relative power, $p$, entering the main spillover region
with respect to the total one, the solid angle subtended by the main spillover 
region, the angle, $\alpha$, between the directions of the main spillover and the
spacecraft spin axis, and, finally, a phase, $\beta$, describing the displacement of the main spillover
centre direction from the plane defined by the spacecraft spin axis and the telescope
line of sight. The first and third of these parameters turn out to be
the most relevant (at least for small $\alpha$, as in the case of {\sc Planck}). 
Also, we have addressed the coupling of this effect
with the observational strategy, with and without displacements
of the spacecraft spin axis from the ecliptical plane. 
We have investigated the relevance of performing multiple surveys and the effect introduced by a possible uncompleteness (or overcompleteness) of one of the 
considered surveys.

The analytical approach, applied to a simple observational strategy
(e.g. the NSS, in the context of {\sc Planck}) and perturbatively in $\alpha$,
has been used to focus on low multipoles, albeit
some general properties have been found for odd and even $\ell$.
The systematic effect vanishes for an even 
number of complete sky surveys, 
except for $\ell=1$ when $\alpha \ne 0$.
We predict a contamination of the dipole itself,
independently of the considered number of surveys and mainly depending
on $p$ and $\alpha$.
We have shown that 
when a phase $\beta$ is introduced,
not only the dipole but also all the odd multipoles
are present to linear order in $\alpha$. 
We find also that the quadrupole is affected and its observed
amplitude is related to the intrinsic sky realization ($T_{2m}^{SKY}$).
A statistical analysis has been performed aimed at the computation
of the probability of increasing or decreasing of the observed
quadrupole amplitude with respect the intrinsic one.

With numerical simulations, based on a dedicated updated code already 
applied in straylight analyses, 
we verify the above results and extend the analysis
to different scanning strategies and significant displacements
of the main spillover centre direction from the spin axis direction
even for higher $\ell$ and uncomplete (or overcomplete) surveys.
Various aspects relevant in CMB space projects
(such as implications for calibration, impact on polarization measurements, 
accuracy requirement in the far beam knowledge for data analysis applications, 
scanning strategy dependence) have been discussed.

Our analysis shows that DSC should be carefully taken into account
for an high precision calibration of the data from the {\sc Planck} receivers
and for an accurate evaluation of the quadrupole amplitude of the CMB pattern.




\bigskip
\bigskip
\noindent
{\bf Acknowledgements.} We warmly acknowledge all the members of the {\sc Planck}
Systematic Effect Working Group for many discussions and collaborations.
It is a pleasure to thank M.~Bersanelli, N.~Mandolesi, P.~Naselsky, F.~Pasian, 
J.~Tauber, and A.~Zacchei for stimulating conversations.
Some of the results in this paper have been derived using {\tt HEALPix}
\citep{gorski05}. We wish to thank the referee for useful comments.
{\it This work has been done in the framework of the
{\sc Planck} LFI activities.}

\end{document}